\PassOptionsToPackage{svgnames}{xcolor}
\documentclass[a4paper,11pt]{article}
\pdfoutput=1

\usepackage[numbers,sort&compress]{natbib}
\usepackage{amsthm, amsmath, amsfonts, amssymb, amscd, mathtools, youngtab, euscript, mathrsfs, verbatim, enumerate, multicol, multirow, bbding, color, babel, esint, geometry, tikz, tikz-cd, tikz-3dplot, array, enumitem, thm-restate, thmtools, datetime, graphicx, tensor, braket, slashed, standalone, pgfplots, ytableau, subfigure, wrapfig, dsfont, setspace, wasysym, pifont, float, rotating, adjustbox, pict2e,array}
\usepackage{amsfonts}
\usepackage{longtable}
\usepackage{tabularx}
\usepackage{caption}
\usepackage{booktabs}
\usepackage{amsmath}
\usepackage{etoolbox}
\usepackage{makecell}
\usepackage{physics}
\usepackage{blindtext}
\usepackage{mathalpha}
\usepackage[all]{xy}

\usepackage[utf8]{inputenc}
\usepackage[normalem]{ulem}
\makeatletter
\numberwithin{figure}{section}
\usetikzlibrary{arrows, positioning, decorations.pathmorphing, decorations.pathreplacing, decorations.markings, matrix, patterns}

\usepackage{physics}
\usepackage{amsmath}
\usepackage{tikz}
\usepackage{mathdots}
\usepackage{yhmath}
\usepackage{cancel}
\usepackage{color}
\usepackage{siunitx}
\usepackage{array}
\usepackage{multirow}
\usepackage{amssymb}
\usepackage{gensymb}
\usepackage{tabularx}
\usepackage{extarrows}
\usepackage{booktabs}
\usetikzlibrary{fadings}
\usetikzlibrary{patterns}
\usetikzlibrary{shadows.blur}
\usetikzlibrary{shapes}

\definecolor{darkgreen}{RGB}{50,150,0}

    {\endtcolorbox}
\usepackage{mdframed}
\mdfdefinestyle{statementstyle}{
   frametitlerule = true,
   frametitlefont = {\normalfont\bfseries},
   frametitlebackgroundcolor = blue!10,
}
\mdtheorem[style=statementstyle]{statement}{Boundary observables in scaling cosmologies}

\usepackage{jheppub} 

\usepackage[T1]{fontenc} 

\title{\boldmath Holographic origin of TCC and the Distance Conjecture}

\author[a]{Alek Bedroya}

\affiliation[a]{Jefferson Physical Laboratory, Harvard University,\\ 17 Oxford St, Cambridge, MA 02138, USA}

\emailAdd{abedroya@g.harvard.edu}

\abstract{One of the unique features of quantum gravity is the lack of local observables and the completeness of boundary observables. We show that the existence of boundary observables for particles with mass $\lim\limits_{t\rightarrow\infty}\frac{m}{H}=\infty$ in scalar field cosmologies where $a(t)\sim t^{p}$ is equivalent to TCC, which implies $p\leq 1$. Moreover, the mass of weakly-coupled particles must decay like $m\lesssim t^{1-2p}$ to ensure that they yield non-trivial boundary observables. This condition can be expressed in terms of the scalar field that drives the cosmology as $m\lesssim\exp(-c\phi)$ where $c$ depends on the scalar potential. The strongest bound we find is achieved for $V\sim \exp(-2\phi/\sqrt{d-2})$ where $c=1/\sqrt{d-2}$. These results connect some of the most phenomenologically interesting Swampland conjectures to the most basic version of holography.}

\begin{document} 
\maketitle
\flushbottom

\tableofcontents
\section{Introduction}

We have learned a lot about the nature of quantum gravity from string theory and our understanding of string theory largely falls into two categories: 1) The non-perturbative data from BPS objects and dualities, and 2) very detailed perturbative understanding of weak coupling limits which lie at the infinite distance limits of the moduli space. The understanding that string theory provides for the weak coupling limits is an important component of what makes string theory what it is. After all, string theory owes its name to these weakly coupled descriptions and finding a fundamental understanding of general features of the infinite distance limits is as deep as understanding why string theory works.

All infinite distance limits of the moduli space in string theory exhibit some universal features which have led to various Swampland conjectures that postulate that these features are fundamental to quantum gravity. Here we focus on two conjectures. 1) The distance conjecture \cite{Ooguri:2006in} which postulates the existence of a tower of light weakly coupled states at every infinite distance limit of the field space with masses that depend exponentially on the distance in the field space. 2) The Trans-Planckian Censorship Conjecture (TCC) \cite{Bedroya:2019snp}  which prohibits the classicalization of Planckian quantum fluctuations and provides a concrete exponentially small upper bound for the scalar potential in the infinite distance limit of the field space. 

Some of the Swampland conjectures have been connected to more fundamental physical principles such as unitarity, causality, or holography \cite{Arkani-Hamed:2021ajd,Hamada:2018dde,Andriolo:2018lvp,Harlow:2018tng}. However, for the distance conjecture and de Sitter conjectures such as TCC, considerable part of the evidence comes from observations in string theory \cite{Obied:2018sgi,Bedroya:2019snp,Andriot:2017jhf,Andriot:2020lea,Andriot:2022bnb,Andriot:2022xjh,Grimm:2018ohb,Corvilain:2018lgw,Grimm:2019ixq} and compatibility with more established Swampland conjectures \cite{Bedroya:2020rac,Bedroya:2020rmd,Ooguri:2018wrx,Mishra:2022fic}\footnote{For ideas regarding the connection between TCC and breakdown of unitarity see \cite{Brandenberger:2021zib,Brahma:2020zpk,Brandenberger:2021pzy}.}. 

In this work, we present an explanation for some aspects of these conjectures based on the holographic principle. What we mean by holographic principle is the most conservative form of holography, which states that physical observables in a gravitational theory must live on the boundary of spacetime. This statement is equivalent to the statement that there are no local physical observables in quantum gravity. One way to see this is that in quantum gravity, we sum over spacetimes of different topologies, and therefore, a unique spacetime manifold does not exist. However, a classical approximation can arise when a particular configuration extremizes the Quantum Gravity path integral. The fact that the classical picture is emergent rather than fundamental is nicely captured by dualities in string theory. For example, consider two T-dual descriptions of a spacetime with one compact dimension of size $R$ in one frame and $l_s^2/R$ in the other frame. As we change $R/l_s$ from very small values to very large values, the semi-classical spacetime that provides the sharpest approximation to the quantum theory transitions from one description to another. There is a mapping between the descriptions, but there is no direct mapping between the spacetimes. For examples, a local wavepacket in the compact dimension in one picture maps to a winding string with no notion of position in the compact dimension. Generally, the notion of spacetime, and any "local" operator associated with the spacetime are expected to be emergent in quantum gravity.

Assuming that the potential always decays exponentially in the asymptotics of field space, we show that there must be universal lower bounds on 1) the decay rate of the potential and 2) the masses of the weakly coupled particles. 

We note that multiple works have argued that the Trans-Planckian problem in cosmology (i.e. violation of the TCC) \cite{Martin:2000xs,Brandenberger:2000wr,Brandenberger:2012aj}, does not lead to inconsistencies at the level of EFT (see \cite{Kaloper:2002cs,Burgess:2020nec,Komissarov:2022gax} for some examples). However, the focus of the Swampland program is precisely the delineation of EFT-consistency from quantum gravity-consistency \cite{Agmon:2022thq}. In this paper, we will pinpoint what goes wrong when TCC is violated in scaling cosmologies. We show that the missing ingredient in the previous analyses was the presence of boundary observables, which is deeply intrinsic to quantum gravity. 

This paper is organized as follows. Before going into details about the effective potential, we first examine its meaning in section \ref{EFT}. We review different definitions for effective action, both based on local observables (e.g. CFT correlators) and boundary observables (e.g. scattering amplitudes). In section \ref{hvp}, we show that a boundary definition for the effective potential is especially needed when the potential is positive. This is mainly due to the fact that any local extremum of the potential will be a de Sitter space with finite spatial volume. Therefore, the convexity theorem applies to any effective potential obtained from the QFT path integral. But as we will discuss, a positive effective potential cannot be everywhere convex. Therefore, a boundary definition of the effective action is required in theories with gravity. 

In section \ref{sec4} we consider a pure quintessence cosmology driven by a scalar field that rolls to infinity in the field space. We study the boundary data that can be produced by weakly coupled particles with masses parametrically heavier than the Hubble energy. Such particles are ubiquitous in the known landscape of the quantum gravity and generically expected \cite{Rudelius:2022gbz}. We show that in some cases, there is no non-trivial boundary data at all. In other words, the field theory would be a uniquely bulk phenomenon, and therefore, it cannot be holographically reproduced. The conditions that we find are
\begin{itemize}
    \item The expansion cannot be accelerated $a\sim t^p,~~p\leq 1$.
    \item For $p>1/2$, the masses of the weakly coupled particles must decay polynomially fast in time $m\lesssim t^{1-2p}$.
\end{itemize}
Note that the results of this section do not apply to cases where the thermal energy-momentum of other fields dominates the evolution (e.g. matter or radiation dominated cosmologies).

In section \ref{sec5}, we discuss the relation between these results and various Swampland Conjectures. The first result is simply equivalent to the TCC at future infinity, which could also be expressed in the language of the de Sitter conjecture. As explained in \cite{Bedroya:2019snp}, TCC implies that the potential must decay as $\exp(-\lambda\phi)$ where $\lambda\geq 2/\sqrt{d-2}$, unless the contribution of an emerging light tower of states to the energy density becomes significant in future infinity. In that case, TCC implies the more conservative bound of $\lambda\geq \frac{2}{\sqrt{(d-1)(d-2)}}$. In \cite{Rudelius:2022gbz}, it was argued that emergent string conjecture \cite{Lee:2019wij} prevents the emergence of such light states in future infinity. In that case, TCC always implies $\lambda\geq 2/\sqrt{d-2}$. 

As for the second condition, when expressed in terms of the scalar field that drives the expansion, it becomes identical to the statement of the distance conjecture $m\lesssim\exp(-c\phi)$. The $c$ that we find depends on $\lambda$. However, the greatest $c(\lambda)$ that we find is $c=1/\sqrt{d-2}$, which remarkably matches with the proposal in \cite{Etheredge:2022opl}, which is proposed on different grounds.

Note that our arguments only apply to universes that have polynomial expansion. Since Hubble scale is constant in de Sitter, and any field with a constant mass $m$ violates our assumption of $\lim\limits_{t\rightarrow\infty}(m/H)=\infty$. Thus, we cannot rule out eternal de Sitter. In fact, an EFT in de Sitter can famously produce boundary data according to dS/CFT \cite{Strominger:2001pn}. It is known that the asymptotic information in universes with an accelerated expansion is spread across different Hubble patches and might not be measurable by a bulk observer \cite{Witten:2001kn,Bousso:2004tv}, which might be a reason to think such universes are not physical \cite{Rudelius:2021azq}. However, our argument points out a specific and fundamental problem with a subclass of universes with accelerated expansion. We argue that boundary observables do not exist for particles parametrically heavier than Hubble, whether we allow meta-observers or not.

\section{Effective action}\label{EFT}

In this section we review two approaches for writing down an effective action for a physical system based on the data used to produce it. Generally, an effective action is the approximation of a quantum physical system by a classical field theory. The physical data that is used to construct the effective action is either the algebra of some local observables, or some boundary observables (e.g. S-matrix). CFTs are examples of the former class that do not have asymptotic states \cite{Weinberg:2012cd,Tanimoto:2013vla}, and quantum gravities are believed to fall in the second category since they do not have gauge-invariant local observables \cite{DeWitt:1967ub}. Note that it is possible to have two dual descriptions of the same theory that fall into different categories. AdS/CFT correspondence \cite{Maldacena:1997re} is an examples of this where one theory is more naturally defined using boundary observables, while the other theory is a CFT that has local observables.   

In the following, we briefly review both of these approaches to the effective action, and study their requirements as well as their physical implications for the underlying physical theory. 
\vspace{5pt}

\textbf{\large{Effective action for local observables}} 

The most common definition of a field theory is a theory with local observables and the effective action captures the local interactions in such a theory. We view a quantum field theory as a theory of local observables defined via path-integral rather than a prescription to calculate scattering amplitudes. 

Suppose we have a field theory with a bare action $S_0$ and a scalar field $\phi(x)$. The effective action $\Gamma(\phi)$ is defined to give us the quantum corrected equations of motion for the expectation values of the operators. We can evaluate the vacuum expectation value of the operator $\hat\phi(x)$ by inserting it into the path integral. If the theory has a Minkowski vacuum, this vev would not depend on $x$. However, we can consider non-vacuum backgrounds for which $\expval{\phi(x)}$ evolves in time. We are particularly interested in classical (or coherent) backgrounds. We can construct a coherent background by acting on the vacuum with $e^{\alpha \hat a-\alpha^*\hat a^\dagger}$, where $\hat a^\dagger$ and $\hat a$ are creation and annihilation operators. In order to act with an operator on vacuum, we can insert it into the path integral. Therefore, to construct a background which looks classical, we can insert an operator $e^{\alpha_p \hat a_p-\alpha_p^*\hat a_p^\dagger}$ for every harmonic oscillator corresponding to a momentum $p$:
\begin{align}
    \int \mathcal{D}\phi e^{iS} e^{\int d^{d-1}p~~\alpha_p \hat a_p-\alpha_p^*\hat a_p^\dagger}.
\end{align}
Given that $a_p$ is the Fourier transform of $\phi(x)$, any linear combinations of $\phi(x)$ takes the form above. Therefore, we can express a coherent state as follows 
\begin{align}
    \int \mathcal{D}\phi e^{iS} e^{i\int d^dx J(x)\phi(x)},
\end{align}
The insertion of $e^{i\int d^dx J(x)\phi(x)}$ creates a classical background and is referred to as the source term. Now that we have created a non-vacuum background, we can re-evaluate the expectation value of our local operator $\hat\phi(x)$ and see how it evolves in time. Let the expectation value of $\phi(x)$ in this background be $\phi_J(x)$. If we take the limit of $\hbar\rightarrow 0$, the quantum fluctuations disappear and the path integral is replaced by its integrand. Therefore, we find that 
  \begin{align}
      \int \mathcal{D}\phi e^{iS} e^{i\int d^dx J(x)\phi(x)}\rightarrow e^{iS(\phi_J)+i\int d^dx J(x)\phi(x)}.
  \end{align}  
In this limit the equations of motion are given exactly by the action $S$. However, when $\hbar$ is non-zero, the above equation breaks. The effective action $\Gamma_{full}$ is defined to capture these quantum effects and maintain the above equation. 
\begin{align}
      \int \mathcal{D}\phi e^{iS+i\int d^dx J(x)\phi(x)}=  \phi e^{i\Gamma_{full}(\phi_J)+i\int d^dx J(x)\phi(x)}.
\end{align}
This definition turns out to capture the fully quantum corrected equations of motion, meaning that the expectation value of local operators follow the equations of motion given by minimizing $\Gamma_{full}$.
\begin{align}\label{eom}
    \frac{\delta\Gamma_{full}(\expval{\phi})}{\delta\expval{\phi(x)}}=0.
\end{align}

If we calculate $\Gamma_{full}$ according to above prescription, it would have all sorts of non-local terms. An important manifestation of the non-local effects is the running of the coupling constants. What we usually mean by an effective action, is an approximation of $\Gamma_{full}$ with a local action $\Gamma$ with finite number of terms. This effective action can then be used to calculate long-range correlation functions. In other words, $\Gamma$ captures the full quantum effect as far as long-range observables are concerned. 

Approximating $\Gamma_{full}$ with a local action $\Gamma$ is not always possible. Even if we start with a local bare action $S$, the quantum effects generate variety of non-local terms in $\Gamma_{full}$. If the field theory is strongly interacting, the non-local terms can be too strong to neglect. Typically, only weakly interacting QFTs can be described an effective action at long ranges. 

Suppose the effective action can be written as a sum of a canonical kinetic term and an effective potential term $V(\phi)$ that only depends on $\expval{\phi}$. In that case the equation \eqref{eom} implies that the vacuum expectation value of $\phi$ must be at a local minimum of $V(\phi)$.
\vspace{5pt}

\textbf{\large{Effective action for boundary observables}}

In quantum gravity, it is often the case that the theory is formulated in terms of some asymptotic observables defined on the conformal boundary of spacetime. The asymptotic data is string theory is the string worldsheet amplitudes, however, the interpretation and the structure governing these amplitudes vary depending on the background. In Minkowski spacetime, the string amplitudes define the S-matrix, while in the AdS spacetime, they have a much richer structure that represent the correlation functions in the dual CFT. The formulation of quantum gravity in terms of boundary data is the general basis for holography. In both Minkowski and AdS backgrounds, we can reproduce the low-energy boundary data of the quantum gravity using a QFT in curved spacetime with an appropriate effective action. This is how one calculates effective action in quantum gravity.

In this section, we review the boundary data in different backgrounds and how they can be calculated using the effective action. If a theory of quantum gravity is formulated based on the corresponding boundary data, one can use the results of this section to reverse engineer an effective action that approximates that theory. 
\vspace{10pt}

\textbf{Minkowski space ($V=0$)}
\vspace{10pt}

The boundary observables in a weakly coupled quantum field theory in Minkowski spacetime are the usual S-matrix elements. In Minkowski spacetime, the LSZ reduction formula allows us to define S-matrix elements in terms of the field theory correlation functions. The resulting S-matrix elements are asymptotic data, as they only depend on the initial condition in the asymptotic past/future\footnote{There are some subtleties with massless particles in 4d due to IR divergences which we will not discuss here. For a discussion of this problem and a proposed resolution see \cite{Hannesdottir:2019opa}.}. For example, for massive scalar particles, the LSZ formula states \cite{Peskin:1995ev}

\begin{align}
&\int \Pi_{i=1}^m\left(d^Dp_i e^{ip_ix_i}\right)\int \Pi_{j=1}^n\left(d^Dq_j e^{-iq_jy_j}\right)\expval{\mathcal{T}\{\phi(x_1)\hdots\phi(x_m)\phi(y_1)\hdots\phi(y_n)\}}\nonumber\\
\sim&~~~\left(\Pi_{i=1}^m\frac{i}{p_i^2-m^2+i\epsilon}\right)\left(\Pi_{j=1}^n\frac{i}{q_j^2-m^2+i\epsilon}\right)\bra{ \bf{p_1}\hdots \bf{p_m}}S\ket{ \bf{q_1}\hdots \bf{q_n}},
\end{align}
where $\mathcal{T}$ is the time ordering operator and $\sim$ is equality up to analytic terms. Note that we have assumed that $\phi$ is canonically normalized which means the kinetic terms of $\phi$ in the effective action takes the form $-\frac{1}{2}(\partial_\mu\phi)^2$. Since the correlation functions can be calculated from the effective action using the Schwinger–Dyson equation, we can also calculate the S-matrix elements from the effective action using tree-level diagrams. 

It is helpful to rewrite the scattering amplitude in the following form for reasons that will become clear shortly.
\begin{align}
    <\tilde\phi_b(P_1)\tilde\phi_b(P_2)...\tilde\phi_b(P_n)>=\lim_{p_i\rightarrow P_i} \Pi_i \tilde G(p_i)^{-1}<\tilde\phi(p_1)\tilde\phi(p_2)...\tilde\phi(p_n)>,
\end{align}
where $P_i$ and $p_i$ are respectively on-shell and off-shell momenta, $G(p)$ is the propagator in the momentum space, and $\phi_b$ are boundary observables. The $P_i$'s with positive energy correspond to in-going states while the ones with negative energy correspond to out going states. 

One of the significant properties of QFT in Minkowski space is that the Hilbert space of asymptotic past/future admits a Fock space representation. As we will see shortly, this is not generically true in an arbitrary spacetime.

\vspace{10pt}

\textbf{Anti-de Sitter space ($V<0$)}
\vspace{10pt}

In AdS space, there is a very rich asymptotic structure given by the rescaled correlation functions at the asymptotic boundary. One can intuitively understand the difference between Minkowski space and AdS space as follows. In AdS space an observer can send a null ray to the boundary and receive its reflection in a finite time. This means, the boundary condition in AdS space is relevant to the experiments inside the AdS. Therefore, there must be a meaningful way of measuring the correlation functions in AdS, even when some of those points approach the conformal boundary. The boundary data for QFT in AdS space is the basic ingredient that makes AdS/CFT correspondence possible. Anti de Sitter space in the global coordinate has the following metric.
\begin{align}
    ds^2=-(1+\frac{r^2}{l_{AdS}^2})dt^2+(1+\frac{r^2}{l_{AdS}^2})^{-1}dr^2+r^2d\Omega_{d-1}^2.
\end{align}
Suppose, $\{{\bf{X}_i}\}_{1\leq i\leq n}$ are $n$ sets of coordinates on the $(D-1)$-sphere. Then, the boundary data is
\begin{align}
    \lim_{r\rightarrow\infty}\expval{\mathcal{T}\{(r^{\Delta_1}\phi_1(t_1,r,{\bf{X_1}})r^{\Delta_2}\phi_2(t_2,r,{\bf{X_2}})\hdots r^{\Delta_n}\phi_n(t_n,r,{\bf{X_n}})\}},
\end{align}
where $\phi_i$ are fields and $\Delta_i$ are appropriate weights that make the above expression convergent. For a free field with mass $m$, $\Delta$ is given by $\Delta=\frac{d}{2}+\frac{1}{2}\sqrt{d^2+4m^2}$\cite{Witten:1998qj}.

The normalized time-ordered correlation functions become singular in the limit where two of the points coincide and the corresponding singularity behaves like an OPE of primary fields in a CFT \cite{Witten:1998qj}. The ADS/CFT correspondence postulates that for any theory of quantum gravity, the normalized time-ordered correlations on the boundary must in fact correspond to the correlation functions of a CFT. Therefore, the asymptotic data of an EFT in AdS is correlation functions. Let us take a moment to explain how these boundary correlation functions are related to string worldsheet amplitudes.

In flat space, the string worldsheet amplitudes have an insertion of the form $\exp(ik.X)$ for any external leg with momentum $k$. Therefore, a generic S-matrix element looks schematically like 
\begin{align}
    A_{\{q_i\}}=\expval{\int \Pi_i dz d\bar z :e^{iq_i\cdot x_i}:\mathcal{V}(z_i,\bar z_i; X,...))}_{worldsheet},
\end{align}
where $:\_:$ is the proper ordering corresponding to the topology of the worldsheet. If we plug this into the LSZ equation and take a Fourier transform of both sides we find
\begin{align}
    \int \Pi_j [d^Dq_j e^{-iq_jx_j}(\frac{i}{q_j^2-m^2+i\epsilon})^{-1}]\expval{\mathcal{T}\{\phi(x_1)\hdots\phi(x_n)\}}\simeq A_{\{q_i\}},
\end{align}
where $x_i$'s are taken to the asymptotic boundary. Let us look at all the ingredients of the above equation and $(\frac{i}{q_j^2-m^2+i\epsilon})^{-1}$ is the inverse of the Green's function to cancel the evolution of $\phi_j(x_j)$ as we take $x_j$ to the asymptotic boundary and ensure that the expression converges to something. In fact, the AdS boundary correlators are also defined using the same prescription. We multiply bulk correlators by the inverse of the Green's function to make sure the evolution of the two cancel each other out as we take the points to the asymptotic boundary. The result of this calculation is parametrized by boundary coordinates. In flat space these are the momenta $q_i$ which pick out a direction, and in AdS they are coordinates $x_i$. This final expression is also the string worldsheet amplitude. In other words, sting theory directly calculates the well-defined boundary observables and effective action is just model that approximates those boundary observables. 

So in general, we have the following prescription to find the asymptotic observables produced by a field theory. If the particles freeze in position space we use
\begin{align}
    \lim_{x_i\rightarrow X_i\in\text{boundary}} \Pi_j G(X_j,x_j)^{-1}\expval{\mathcal{T}\{\phi(x_1)\hdots\phi(x_n)\}}.
\end{align}
and if they freeze in the momentum space, we use
\begin{align}
       \lim_{p_i\rightarrow P_i: \text{on-shell}} \Pi_i \tilde G(p_i)^{-1}<\tilde\phi(p_1)\tilde\phi(p_2)...\tilde\phi(p_n)>.
\end{align}
\vspace{5pt}

\section{\mbox{Why gravity needs boundary observables when $V>0$?}}\label{hvp}

In this section we want to point out a limitation to the observable-based approach when the scalar potential is positive. Let us first see what happens when the spatial volume is finite. The effective action captures the equations of motion of local quantum observables. So one might think that the lack of need of asymptotic states will put finite and infinite spacetimes on equal footings. However, it turns out that there is a major distinction between the two cases. In finite spacetimes, one can go from any field configuration $\phi(x)=\phi_1$ to any other field configuration $\phi(x)=\phi_2$ with finite action. In other words, the amplitude of tunneling between the two is finite. This is in sharp contrast with infinite spacetimes with dimensions $d>2$ where it takes infinite energy to change the boundary conditions. For this reason, infinite spacetimes can have different boundary conditions parameterized by moduli, that set the theory in different vacua. But in finite spacetimes, the moduli cannot be frozen, as quantum fluctuations can take the system from any field configuration to any other configuration with finite amplitude. 

In the presence of different vacua, the effective potential can have multiple minima each corresponding to a different vacuum. Note that all the minima must have the same energy, otherwise, we can tunnel from one to another using a Coleman instanton in finite time \cite{Coleman:1977py,Coleman:1978ae}. Given that the effective action represents the fully quantum corrected equations, being in the local minima of the effective action must mean that we are in the true vacuum. Therefore, the local minima of the effective action must be impervious to Coleman instantons. Thus, the local minima of the effective potential, must be the global minima of the effective potential. Note that in the case of $V>0$, a tunneling of a growing region of spacetime to a the new vacuum suffices for this argument, even if the expansion of the inter-bubble space keeps the growing bubbles from filling the spacetime.

As we discussed above, infinite spacetimes with $d>2$ dimensions can have multiple vacua. Therefore, the effective potential in infinite spacetimes can have multiple global minima. This is because, the energy cost to transition between the two is infinite and therefore, the amplitude is zero. 

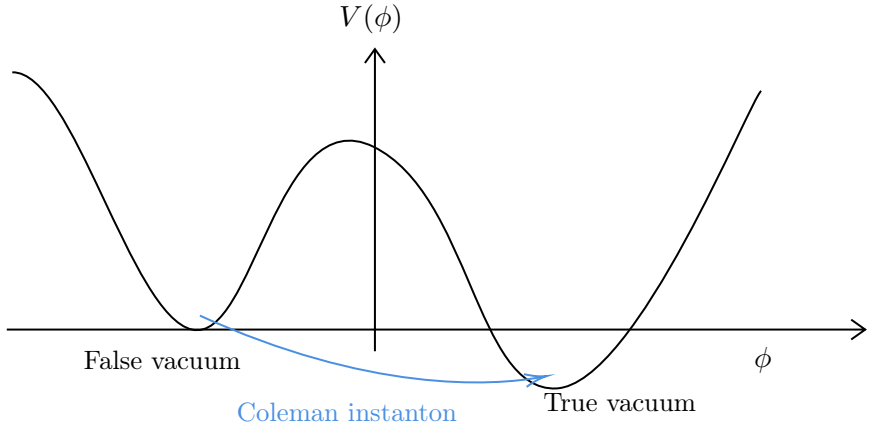
\begin{figure}[H]
    \centering

\tikzset{every picture/.style={line width=0.75pt}} 

\begin{tikzpicture}[x=0.75pt,y=0.75pt,yscale=-1,xscale=1]

\draw  (138,171.24) -- (566.5,171.24)(321.76,30.16) -- (321.76,182.15) (559.5,166.24) -- (566.5,171.24) -- (559.5,176.24) (316.76,37.16) -- (321.76,30.16) -- (326.76,37.16)  ;
\draw    (140.75,41.85) .. controls (171.05,41.07) and (199.78,164.22) .. (230.47,171.24) .. controls (261.16,178.25) and (272.79,55.54) .. (320.97,79.26) .. controls (369.15,102.99) and (372.12,202.7) .. (412.26,200.85) .. controls (452.39,199) and (509.09,55.27) .. (514.56,51.21) ;
\draw [color={rgb, 255:red, 74; green, 144; blue, 226 }  ,draw opacity=1 ]   (234.4,164.22) .. controls (317.69,202.41) and (379.07,200.18) .. (405.93,194.94) ;
\draw [shift={(407.53,194.62)}, rotate = 168.14] [color={rgb, 255:red, 74; green, 144; blue, 226 }  ,draw opacity=1 ][line width=0.75]    (10.93,-3.29) .. controls (6.95,-1.4) and (3.31,-0.3) .. (0,0) .. controls (3.31,0.3) and (6.95,1.4) .. (10.93,3.29)   ;

\draw (302.47,6.08) node [anchor=north west][inner sep=0.75pt]    {$V( \phi )$};
\draw (509.53,178.33) node [anchor=north west][inner sep=0.75pt]    {$\phi $};
\draw (251.32,206.42) node [anchor=north west][inner sep=0.75pt]  [font=\small,color={rgb, 255:red, 74; green, 144; blue, 226 }  ,opacity=1 ] [align=left] {Coleman instanton};
\draw (174.99,180.27) node [anchor=north west][inner sep=0.75pt]  [font=\small] [align=left] {False vacuum};
\draw (404.35,201.32) node [anchor=north west][inner sep=0.75pt]  [font=\small] [align=left] {True vacuum};

\end{tikzpicture}
    \caption{If the potential has another local minimum with a smaller energy, due to Coleman instantons, the universe will nucleate expanding bubbles of the true vacuum which collectively fill the universe in finite time.}
    \label{Coleman}
\end{figure}
It is often said that the effective action is convex, however, the convexity theorem only applies when one has not specified the boundary conditions of path integral at infinity. When one does not specify that boundary condition, they are effectively averaging over all different vacua. But, when fixing the boundary condition at infinity, one can derive a non-convex effective action that captures the fully quantum equations of motion. Higgs potential is an example of this.

\begin{figure}[H]
    \centering

\tikzset{every picture/.style={line width=0.75pt}} 

\begin{tikzpicture}[x=0.75pt,y=0.75pt,yscale=-1,xscale=1]

\draw  (126,183.18) -- (548.5,183.18)(307.18,30.57) -- (307.18,194.99) (541.5,178.18) -- (548.5,183.18) -- (541.5,188.18) (302.18,37.57) -- (307.18,30.57) -- (312.18,37.57)  ;
\draw    (128.72,43.22) .. controls (158.59,42.37) and (186.91,175.59) .. (217.17,183.18) .. controls (247.43,190.77) and (264.51,72.73) .. (306.41,83.69) .. controls (348.31,94.65) and (358.39,185.18) .. (397.97,183.18) .. controls (437.54,181.18) and (491.9,57.73) .. (497.29,53.33) ;
\draw    (347.14,225.34) .. controls (377.94,226.17) and (391.22,220.51) .. (398.32,194.1) ;
\draw [shift={(398.74,192.46)}, rotate = 104.09] [color={rgb, 255:red, 0; green, 0; blue, 0 }  ][line width=0.75]    (10.93,-3.29) .. controls (6.95,-1.4) and (3.31,-0.3) .. (0,0) .. controls (3.31,0.3) and (6.95,1.4) .. (10.93,3.29)   ;
\draw    (276.14,224.5) .. controls (232.27,226.96) and (218.63,216.6) .. (218.69,191.86) ;
\draw [shift={(218.72,189.93)}, rotate = 91.7] [color={rgb, 255:red, 0; green, 0; blue, 0 }  ][line width=0.75]    (10.93,-3.29) .. controls (6.95,-1.4) and (3.31,-0.3) .. (0,0) .. controls (3.31,0.3) and (6.95,1.4) .. (10.93,3.29)   ;

\draw (276.23,5.03) node [anchor=north west][inner sep=0.75pt]    {$V_{\{eff\}}( \phi )$};
\draw (492.23,191.52) node [anchor=north west][inner sep=0.75pt]    {$\phi $};
\draw (274.79,216.42) node [anchor=north west][inner sep=0.75pt]  [font=\small] [align=left] {True vacua};

\end{tikzpicture}
    \caption{In infinite spacetimes with dimension $d>2$, there could be different vacua. All of these vacua must have the same energy to ensure their stable against Coleman instantons. Moreover, a true vacuum must be the minimum of the effective potential. Therefore, an effective potential that is calculated via fixing the boundary condition at infinity, can have multiple minima, and is not necessarily convex.}
    \label{Multiple vacua}
\end{figure}
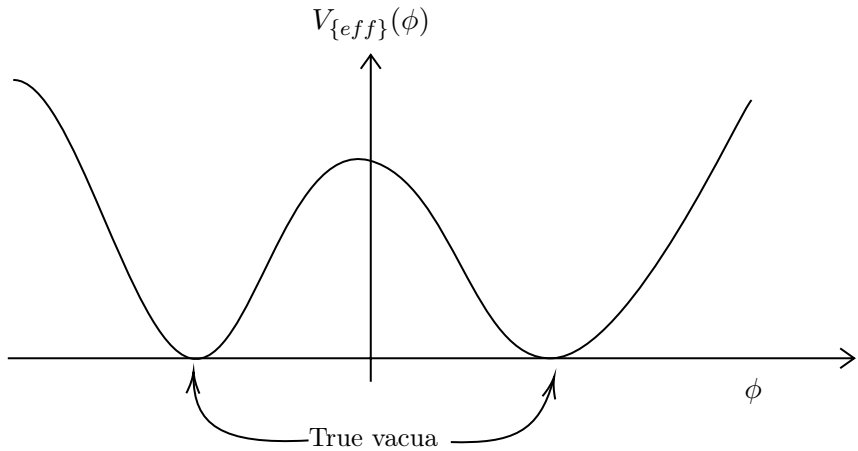

Going back to case of finite volume, no two minima are separated vacua since we can always transition between states with finite amplitude. Therefore, there must be a unique vacuum in the theory. This implies that the effective potential must have a unique global minimum. This in fact follows from the convexity theorem as well \cite{Peskin:1995ev}. In finite volume there are no boundary conditions to fix, and therefore the convexity theorem applies. A convex function has a unique minimum, therefore, the vacuum is unique. de Sitter space has a finite volume, therefore, if it is realized as a minimum of an effective potential which is defined based on the local observable approach, that potential must be convex. However, in all known examples in string theory the potential dies off exponentially in all directions of the field space \cite{Obied:2018sgi} and the following three criteria are mutually inconsistent:
\begin{itemize}
    \item $V$ is convex.
    \item $V$ decays exponentially at infinities. 
    \item $V$ has an extremum with $V>0$.
\end{itemize}
In fact, as long as $V>0$ and decays exponentially at infinities, it will always have an extremum. Thus, we do not even need to assume the existence of an unstable or metastable de Sitter as long as $V>0$. 

The above argument shows that the effective action based on local observables especially does not make sense when $V>0$ and at best, it can only explain a finite window of field space. Therefore, to have a meaningful definition of effective potential, gravity necessitates existence of boundary observables for $V>0$.


\section{Boundary observables in scaling cosmologies}\label{sec4}

In this section we study the boundary observables that can be defined in FRW backgrounds with polynomial expansions $a\sim t^p$. The reason we are interested in these backgrounds is that as we show in the Appendix \ref{A1} any expansion driven by an exponentially decaying potential is polynomial\footnote{Most of the calculations in the Appendix \ref{A1} are standard calculations that we include for the sake of completeness (see for example \cite{Rudelius:2021azq,Hellerman:2006nx,Bedroya:2020rmd}).}. Exponential potentials are ubiquitous in string theory and in all the known examples, the scalar potentials decays exponentially at the infinity of the filed space. This also follows from the Swampland de Sitter conjecture \cite{Obied:2018sgi,Garg:2018reu,Ooguri:2018wrx}. 

\subsection{Asymptotic cosmology for exponential potentials}

Let us summarize the results of the Appendix \ref{A1}. Assuming that the evolution of spacetime is driven by an exponentially decaying scalar potential, we find
\begin{itemize}
    \item Expansionary/contracting solutions expand/contract polynomially in the future/past infinity.
    \item Unless there is a bounce, every spacetime has a spacelike singularity at finite time (i.e. big bang or big crunch).
    \item Suppose the scale factor goes like $a\sim t^p$ at the asymptotic of the field space and scalar potential goes like $\exp(-\lambda\phi)$. We have 
    \begin{align}
        &\lambda <2\sqrt{\frac{d-1}{d-2}}: p=\frac{4}{(d-2)\lambda^2}\nonumber\\
        &\lambda >2\sqrt{\frac{d-1}{d-2}}: p=\frac{1}{d-1}.
    \end{align}
    \item $p>1$ iff $\lambda<\frac{2}{\sqrt{d-2}}$.
\end{itemize}

The above results show the connection between exponential potentials and polynomial evolutions. In the rest of the section, we focus on the cosmological evolution and study the boundary observables for a given background that expands polynomially as
\begin{align}
    a(t)\sim t^p,
\end{align}
as $t\rightarrow\infty$ where $a(t)$ is the scale factor in the FRW solution with flat spatial curvature
\begin{align}
    ds^2=-dt^2+a(t)^2(\sum _{i=1}^{d-1}dx^{i~2}).
\end{align}

\subsection{Heuristic calculations}

Before doing any precise calculations, let us do some heuristic calculations to see what we should expect about the boundary data. Consider a particle with mass $m$ and comoving momentum $k$. The proper momentum of the particle is $k/a$. Therefore, at late times, we expect the particle to have energy $\omega\simeq m+\frac{k^2}{2ma^2}$ and comoving velocity $v\simeq \frac{k}{ma^2}$. The maximum distance that this particle can travel in comoving coordinate is
\begin{align}
    \int dt\frac{k}{ma^2},
\end{align}
which is finite if $p>1/2$. Even if $m$ depends on the running scalar and changes with time as $t^{-q}$, as long as $q<2p-1$, the particle cannot make it to spatial infinity and freezes out at some comoving coordinates. Therefore, we cannot define an S-matrix and our boundary data must be frozen correlators. 

Now let us consider the massless case. For massless particles, the particle horizon is given by 
\begin{align}
    \int \frac{dt}{a}.
\end{align}
Therefore, if $p>1$, the particle massless particle freezes out at some comoving coordinate and our boundary data must be frozen correlators rather than scattering amplitudes.

In the following subsections, we see how the boundary correlation functions freeze when $q<\min\left(2p-1,1\right)$. Moreover, we show that all such vanish when the boundary points are not coincident.

\subsection{Boundary correlation functions}

The Ricci curvature in this background is
\begin{align}\label{RFRW}
    \mathcal{R}=2(d-1)\frac{\ddot a}{a^2}+(d-1)(d-2)\frac{\dot a^2}{a^2}.
\end{align}
In a general curved background, the equation of motion of a scalar field is given by 
\begin{align}
    [-\frac{1}{\sqrt{-g}}\partial_\mu(g^{\mu\nu}\sqrt{-g}\partial_\nu)+m^2+\xi\mathcal{R}]\phi=0,
\end{align}
where $m$ is the mass and $\xi$ is the coefficient of the $\phi\mathcal{R}$ coupling. We assume $\xi=0$ as its value does not affect our conclusions. 
\begin{align}\label{eom0}
    \frac{1}{a^2}[-\Delta^2+(d-1)\dot a a\partial_t+a^2\partial_t^2+m^2 a^2]\phi=0.
\end{align}
Since we have translational symmetry in space, we can consider the following solution ansatz.
\begin{align}
    \tilde\phi_k=e^{-i\omega(t)\cdot t+i\vec k\cdot \vec x}.
\end{align}
Note that $\omega$ depends on $t$. As $t\rightarrow\infty$, the last two terms dominate and we find
\begin{align}
    \lim_{t\rightarrow\infty} \omega(t)\rightarrow \pm m.
\end{align}

Let us focus on the positive frequency modes for now. We define $\omega=m+\delta\omega$, where $\delta\omega$ vanishes at $t=\infty$.

\begin{align}
    \frac{1}{a^2}[k^2-i(d-1)\dot a a(\omega+t\dot \omega)-(\omega+t\dot \omega)^2a^2+m^2a^2]\tilde\phi_k=0.
\end{align}
For $\omega$ to converge we need $t\dot\omega\rightarrow 0$. Therefore, at late times, we can replace $\dot a a(\omega+t\dot \omega)$ with $\dot a am$. However, we cannot do the same with  $(\omega+t\dot \omega)^2a^2$. This is because $t\dot\omega\omega a^2$, $2\omega\delta\omega a^2$ could be relevant. However, the rest of the terms will be subleading. So we find
\begin{align}
    \frac{1}{a^2}[k^2-i(d-1)\dot a a m-2m\delta\omega a^2-2t\delta\dot\omega m a^2]=0,
\end{align}
which leads to
\begin{align}
    t\dot{(\delta\omega)}+\delta\omega+i\frac{d-1}{2}\cdot\frac{\dot a}{a}-\frac{k^2}{2ma^2}=0.
\end{align}
The general solution to the above differential equation is
\begin{align}
    \delta\omega=\frac{c}{t}-i\frac{(d-1)p}{2}\frac{\ln(t)}{t}-\frac{k^2}{2ma^2(2p-1)},
\end{align}
where $c$ is a constant. The first term amounts to an overall normalization of $\tilde\phi$. We choose $c$ such that
\begin{align}
\delta\omega=-i\frac{1}{t}\ln((\frac{a}{a_0})^{(d-1)/2})-\frac{k^2}{2ma^2(2p-1)}.
\end{align}
For $\omega=m+\delta\omega$ we find 
\begin{align}
    \omega=m-i\frac{1}{t}\ln((\frac{a}{a_0})^{(d-1)/2})-\frac{k^2}{2ma^2(2p-1)}.
\end{align}
As we states above, there is also a negative frequency mode which we can obtain by fliping the sign of $m$ in the above equation. So we find
\begin{align}\label{abo}
     \omega_\pm(k,t)=\pm m-i\frac{1}{t}\ln((\frac{a}{a_0})^{(d-1)/2})\mp\frac{k^2}{2ma^2(2p-1)}.
\end{align}
We could relate the two solutions to each other by
\begin{align}\label{pmr}
    i\omega_\pm=(i\omega_\mp)^*.
\end{align}
This follows from the fact that if $\tilde\phi$ is a solution so must be $\tilde\phi^*$ because the equations are real. This implies that if $\omega$ gives a solution, so does $-\omega^*$. Therefore, either $\omega$ is purely imaginary or the two frequencies are related by the above relation. 

We consider $p>1/2$ where the friction term in \eqref{abo} dominates and the modes freeze at time infinity.

Now let us quantize our scalar field. The canonical quantization relations lead to 
\begin{align}\label{cqr}
    [\phi(x),\pi(y)]=i\delta^{d-1}(x-y)a^{-(d-1)},
\end{align}
where $\pi=\partial_t\phi$. The factor of $a^{-(d-1)}$ shows up because the distance $x-y$ in the delta function is not the proper distance. We could equivalently write 
\begin{align}
    [\phi(x),\pi(y)]=i\delta^{d-1}(a(x-y)).
\end{align}
Note that the above equation uses the fact that metric is diagonal and $g_{tt}=-1$. Otherwise, we would have $\pi=-\partial^t\phi$ which is not necessarily the same as $\partial_t\phi$.

Now let us expand the operators $\phi$ and $\pi$ in terms of the solutions $\tilde\phi_k$. 
\begin{align}
    \phi(x,t)=\int d^{d-1}k~a_k e^{-i\omega_+(k,t)t+i\vec k\cdot\vec x}+a_k^\dagger e^{-i\omega_-(k,t)t-i\vec k\cdot\vec x}.
\end{align}
The reality of the above expression follows from \eqref{pmr}. Then we have
\begin{align}
    \pi(y,t)=\int d^{d-1}k'~a_{k'}(-i)(\omega_++\dot\omega_+t) e^{-i\omega_+(k',t)+i\vec k'\cdot\vec y}+a_{k'}^\dagger (-i)(\omega_-+\dot\omega_-t)e^{-i\omega_-(k',t)-i\vec k'\cdot\vec y}.
\end{align}
From the above equations we find
\begin{align}\label{cre}
    [\phi(x,t),\pi(y,t)]=\int d^{d-1}k~d^{d-1}k'&[a_k,a_{k'}](-i)(\omega_++\dot\omega_+t)e^{-i\omega_+(k,t)t-i\omega_+(k',t)t+i\vec k\cdot\vec x+i\vec k'\cdot\vec x}\nonumber\\
    +&[a_k^\dagger,a_{k'}^\dagger](-i)(\omega_-+\dot\omega_-t)e^{-i\omega_-(k,t)t-i\omega_-(k',t)t-i\vec k\cdot\vec x-i\vec k'\cdot\vec x}\nonumber\\
    +&[a_k,a_{k'}^\dagger](-i)(\omega_-+\dot\omega_-t)e^{-i\omega_+(k,t)t-i\omega_-(k',t)t+i\vec k\cdot\vec x-i\vec k'\cdot\vec x}\nonumber\\
    +&[a_k^\dagger,a_{k'}](-i)(\omega_++\dot\omega_+t)e^{-i\omega_-(k,t)t-i\omega_+(k',t)t-i\vec k\cdot\vec x+i\vec k'\cdot\vec x}.
\end{align}
From \eqref{cqr} we know that the right hand side must depend on time like $a^{-(d-1)}$. If we plug in the asymptotic behavior of $\omega_{\pm}$ from \eqref{abo}, we will see that the only terms that reproduce that asymptotic behavior are coefficients of $[a_k,a_{k'}^\dagger]$ and $[a_k^\dagger,a_{k'}]$. Therefore, we find
\begin{align}
    &[a_k,a_{k'}]=0,\nonumber\\
    &[a_k^\dagger,a_{k'}^\dagger]=0.
\end{align}
We can rewrite the rest of the expression \eqref{cre} follows.
\begin{align}
      [\phi(x,t),\pi(y,t)]=\int d^{d-1}k~d^{d-1}k'[a_k,a_{k'}^\dagger](-2i)(\omega_-+\dot\omega_-t)e^{-i\omega_-(k,t)t-i\omega_-(k',t)-i\vec k\cdot\vec x-i\vec k'\cdot\vec x}.
\end{align}
Using \eqref{abo}, at $t\rightarrow\infty$, we find
\begin{align}
     [\phi(x,t),\pi(y,t)]\simeq \int d^{d-1}k~d^{d-1}k'(2im)(\frac{a_0}{a})^{d-1}[a_k,a_{k'}^\dagger].
\end{align}
For \eqref{cre} to be true, we need
\begin{align}\label{cra}
    [a_k,a_{k'}^\dagger]=\frac{\delta^{d-1}(k-k')}{2ma_0^{d-1}}.
\end{align}
Now let us define the vacuum $\ket{\Omega}$ to be the state that is annihilated by all the operators $a_k$. This is a natural choice to ensure that at small scales, the vacuum looks like Minkowski (i.e. equivalence principle). Since any comoving mode is trans-Planckian at some time in the past, if it is not in the vacuum, we will have a state that has highly UV excitations. Therefore, if the emerging modes are not in the vacuum, then we have an immediate Trans-Planckian problem. 

Let us calculate the equal-time correlation function for this state. When we expand the fields in terms of the creation and annihilation operators, there is only one combination that does not vanish. 

\begin{align}
    \bra{\Omega}\phi(x,t)\phi(y,t)\ket{\Omega}=\int d^{d-1}k~d^{d-1}k' \bra{\Omega}a_k a_{k'}^\dagger\ket{\Omega} e^{-i\omega_+(k,t)t-i\omega_-(k',t)+i\vec k\cdot \vec x-i\vec k'\cdot\vec y}.
\end{align}
Using $\bra{\Omega}a_k a_{k'}^\dagger\ket{\Omega}=\bra{\Omega}[a_k, a_{k'}^\dagger]\ket{\Omega}$ and the commutation relation \eqref{cra}, we find
\begin{align}\label{ID}
    \phi(x,t)\phi(y,t)=\int d^{d-1}k~\frac{1}{2ma_0^{d-1}}e^{-2i\text{Im}(\omega(k,t)t)+i\vec k\cdot (\vec x-\vec y)}.
\end{align}
We have dropped the state $\ket{\Omega}$ for simplicity. Since the leading $k$-dependence of $\omega$ in \eqref{abo}, the above expression gives a simple delta function at $t\rightarrow\infty$.
\begin{align}\label{TBC}
    \phi(x,t)\phi(y,t)\simeq \frac{1}{2ma^{d-1}}\delta^{d-1}(x-y).
\end{align}
We can use factor out the asymptotic time-dependence of $\omega$ by defining boundary fields as $\phi_b(x)\equiv \lim_{t\rightarrow\infty}\phi(x,t)a(t)^{(d-1)/2}$. We find that the boundary correlators are given by 
\begin{align}
    \phi_b(x)\phi_b(y)=\frac{1}{2m}\delta^{d-1}(x-y),
\end{align}
and so are trivial. 

We can use adiabatic approximation to see what happens if $m$ depends on the running scalar and therefore changes with time. Assuming $m$ depends polynomially on time, as long as $m$ decays slower than $H\sim 1/t$ and $1/(Ha^2)\sim t/a^2$, the mass term in in \eqref{abo} is leading and the dominant sub-leading correction is k-independent. For such particles, the above argument could be repeated to deduce that the correct asymptotic data must be frozen correlators which happen to be trivial. Therefore, as long as $$\lim\limits_{t\rightarrow\infty}\max\left(\frac{m}{H},\frac{ma^2H}{m_{\rm pl}^2}\right)=\infty\,,$$ there are no non-trivial boundary data associated with the field. Note that among the two terms on the LHS, for $p>1$ $\frac{m}{H}$ dominates and for $p<1$ $\frac{ma^2H}{m_{\rm pl}^2}$ dominates.

Let us summarize the results of this section.
 
\begin{statement*}
Suppose we have an FRW background that expands as $a\sim t^p$ in the future infinity. If $p>1$, no massive field satisfying $\lim\limits_{t\rightarrow\infty}\frac{m}{H}=\infty$  yields any non-trivial boundary observable. If $p<1$, a weakly coupled massive particle can produce a non-trivial boundary observable if and only if $m\lesssim t^{1-2p}$ as $t\rightarrow\infty$. 
\end{statement*}

Note that in scaling cosmologies, we expect to have a tower of weakly coupled particles in the asymptotic future whose mass scale is not parametrically smaller than the Hubble parameter and is typically parametrically greater \cite{Rudelius:2022gbz}. Therefore, the assumption $\lim\limits_{t\rightarrow\infty}\frac{m}{H}=\infty$ is generically satisfied in the quantum gravity landscape. 

\subsection{Some important remarks}

\begin{itemize}
\item \textbf{Higher spins: }The calculation of the previous subsection is qualitatively similar for particles of higher spins and the same results apply.

\item \textbf{Weak coupling: }Our results only apply to particles hat become weakly coupled in the future infinity. There could be more massive states that cannot be understood as sharp resonances and therefore, do not lead to a boundary field. For example in a scattering theory, such states cannot be in/out going states because they are not stable. However, they leave an imprint on the scattering amplitude. 

\item \textbf{Meaning of trivial boundary correlators: }
Let us point out that the boundary correlators could be different if they are evaluated for a non-vacuum state in the bulk. In dS/CFT this corresponds to having some operator insertions at the past boundary, which can be expressed as operator insertions at the antipodal point in the future boundary. On the other hand, in AdS/CFT, this corresponds to evaluating the boundary correlators at a state other than the AdS vacuum. In that case, we can create the other state by inserting some other operator on the boundary and the said correlator is equivalent to a more complicated vacuum correlation function. This is a consequence of crossing symmetry or CPT and the same thing can be done here. We can study the two-point function in a non-vacuum background by inserting other operators that change the state from vacuum. For example, $\int d^{d-1}z f(z)\phi_b(z)$ corresponds to a one-particle state and sandwiching the two-point function between this operator corresponds to the two-point function in presence of a particle. 
\begin{align}
    \expval{\phi_b(X)\phi_b(Y)}_{\text{1-particle state}}=\expval{(\int d^{d-1}z' f(z')^*\phi_b(z'))\phi_b(X)\phi_b(Y)(\int d^{d-1}z f(z)\phi_b(z))}_{\ket\Omega}.
\end{align}
As one can see from the above equation, such two-point functions can be reduced to higher-order correlation functions. However, when the two-point functions are delta functions, the correlation functions factorize and can be calculated using Wick's theorem. For example, for the above correlation function we have
\begin{align}
      \expval{\phi_b(X)\phi_b(Y)}_{\text{1-particle state}}\propto f(X)^*f(Y)+f(Y)^*f(X).
\end{align}
Therefore, there is no additional data to the boundary observables and the boundary data is trivial.  In other words, the prefactors of the delta functions (e.g. the factor of $1/2m$ in \eqref{TBC}) are not sufficient to capture the interactions in the bulk and reconstruct the EFT.

Intrestingly, the factorization phenomenon seems to be a universal feature of field theories at infinite distance limits where the theory becomes weakly coupled \cite{Stout:2022phm}. Therefore, it should be expected that frozen correlation functions will factorize and not be able to capture the interactions away from the infinite distance limit of the field space.

\item \textbf{Uniqueness of boundary observables: }One might ask why did we take the boundary limit as $t\rightarrow\infty$ while keeping the spatial co-moving coordinates fixed? For example, if we had taken the proper distance to be fixed rather than the comoving distance, then we would have gotten a different result. However, we argue that there is a unique good choice which is forced on us by the background. Suppose we had chosen to define the boundary propagators as $\phi_b(X)=\lim_{t\rightarrow\infty} f(t)\phi(X t^\alpha)$ for some alpha, where $f(t)$ is chosen such that the limit of boundary correlators exist. If $\alpha<0$, then we are measuring correlators at distances that are shrinking in comoving coordinates. Therefore, such a correlator is set by the modes with decreasing comoving wavelength. However, since these modes were Trans-Planckian at some point, they must be set in the vacuum. In other words, even though the correlation functions are non-trivial, the information of an initial condition is completely lost. On the other hand if $\alpha>0$, the boundary correlator will be even more trivial as it will be identically zero and we will lose the singularity. In general, we must choose the boundary coordinates such that we have a well-defined quasi-stationary mode expansion with constant wavenumbers. 

\item \textbf{Significance of polynomial expansion: }Note that the polynomial dependence on time played an important role in the analysis of this section. The future boundary of universes with exponential expansion or polynomial accelerated expansion are both spacelike. However, in de Sitter space where the expansion is exponential, one can define non-trivial boundary correlators. Since the points on the future boundary are not causually related, the boundary correlators correspond to non-vanishing superhorizon correlations.  These correlators are the basis for dS/CFT \cite{Strominger:2001pn}. The key difference is that when the expansion is exponential, the Hubble parameter stays constant. Therefore, the mass of the particles does not get infinitely large compared to the Hubble energy. Now let us give a more precise explanation and point out how the calculations are different between the exponential and polynomial expansions. In exponential expansion, the friction term in the equation of motion \eqref{eom0} is of the same order as the mass term and does not die off with time. The friction term gives an imaginary component to $\omega(t\rightarrow\infty)$ which then leads to imaginary dependence on $k$ in \eqref{ID}. Therefore, the boundary propagator has a non-vanishing k-dependence as $t\rightarrow\infty$.

In \cite{Dijkgraaf:2016lym}, some exotic string theory backgrounds using negative branes where studied. If such theories exist, the worldvolume theory of the negative branes would be a non-unitary supergroup gauge theory and their dual near horizon geometry are dS and AdS spaces of exotic signatures. There are still some open questions about these theories, in particular, an independent path integral definition of the worldvolume theory of the branes is missing. However, there is some evidence from dualities and string worldsheet for their existence \cite{Dijkgraaf:2016lym}. What is interesting is that all of the exotic string theory backgrounds that where find in \cite{Dijkgraaf:2016lym} have an exponential expansion of the induced volume in the stationary coordinate, and therefore, cannot be ruled out by our argument. 
\end{itemize}

\section{Swampland conjectures from holography}\label{sec5}

In section \ref{hvp} we showed that in quantum gravity, a global positive potential requires a boundary observables to be well-defined which is the most conservative form of holography. In the previous section, we showed that depending on the masses of the particles and the rate of expansion, such boundary condition might not exist. In this section we show that these conditions match with some of the Swampland constraints and can be viewed as a holographic derivation for them.

\subsection{Trans-Planckian Censorship Conjecture}\label{TCC}

Trans-Planckian Censorship Conjecture (TCC) states that in a theory of quantum gravity, we cannot have an expansion that is fast and long enough to stretch Planckian modes beyond the Hubble scale \cite{Bedroya:2019snp}. In other words, the scale factor and Hubble parameter must satisfy the following condition for any given times $t_i$ and $t_f$.
\begin{align}
    \frac{a(t_f)}{a_(t_i)}<\frac{1}{H(t_f)}.
\end{align}
The initial motivation of TCC was that such expansions erase the quantum information as they classicalize the initial quantum fluctuations. The motivations for this conjecture were expanded in a subsequent work \cite{Bedroya:2020rmd}. For backgrounds with polynomial expansion $a\sim t^p$, TCC is equivalent to $p\leq 1$. Therefore, the bound we found for the exponent of the expansion, is equivalent to the statement of TCC in the asymptotic regions of the field space. This is very much in the spirit of the initial motivation of TCC that information somehow gets lost if TCC is violated. As we can see now that the information (whether quantum or classical) completely gets lost at future infinity if TCC is violated.

\subsection{de Sitter conjecture}

We can express the statement of TCC in the asymptotic region of field space in terms of the scalar potential. Using the results of Appendix \ref{A1}, we can see that $p\leq 1$ is equivalent to $\lambda\geq 2/\sqrt{d-2}$ where $V\sim\exp(-\lambda\phi)$. However, as was initially pointed out in \cite{Bedroya:2019snp}, this is only true if we assume that the scalar fields are the main driver of the expansion. One could imagine a scenario where the tower of light states that emerge in the infinite distance limit are excited and contribute to the expansion. In that case the constraint on the potential would be milder and it would be $\lambda\geq2/\sqrt{(d-1)(d-2)}$ \cite{Bedroya:2019snp}.

In \cite{Rudelius:2022gbz}, using the emergent string conjecture, it was argued that the tower of light states are always heavier than the Hubble scale in which case they would not significantly contribute to the expansion. In that case, TCC gives $\lambda\geq 2/\sqrt{d-2}$.

\subsection{Distance conjecture}

In the previous section, we saw that if the mass of the states in the bulk do not go to zero fast enough, they will not lead to any boundary observables. Assuming that the expansion is driven by a potential $V\sim\exp(-\lambda\phi)$, we can express the conditions imposed on their masses as $m\lesssim\exp(-c(\lambda)\phi)$ where $c$ would depend on $\lambda$. For $\lambda<\frac{2}{\sqrt{d-2}}$ no state yields boundary observables and for $\lambda>\frac{8}{\sqrt{(d-2)}}$, any mass yields scattering amplitude. However, for a $\lambda$ in the above range, using the results of Appendix \ref{A1}, our conditions lead to the following non-trivial bound.
\begin{align}
m\lesssim \exp(-c(\lambda)\phi),~~~
    c(\lambda)=\frac{4}{(d-2)\lambda}-\frac{\lambda}{2}.
\end{align}
Note that $c(\lambda)$ is greatest at $\lambda=2/\sqrt{d-2}$ for which $c=1/\sqrt{d-2}$. This is a derivation of distance conjecture in scaling cosmologies. What is even more interesting, is that the coefficient $c=1/\sqrt{d-2}$  matched with the proposal in \cite{Etheredge:2022opl} to sharpen the distance conjecture. However, we arrive at these decay rates from a completely different perspective based on the most basic form of holography.

\subsection{Generalized distance conjecture}

The bound $m\lesssim t^{1-2p}$ can be expressed in terms of the Hubble parameter as $m\lesssim H^{2p-1}$. We can understand this bound in the following way: as we change the geometry of the spacetime in the future infinity, there must be a tower of light states whose mass depends polynomially on the energy density. This statement is similar to the distance conjecture when we view the background metric as a modulus and we think about moving in the space of geometries as going to the asymptotic direction in the moduli space\footnote{We are thankful to Cumrun Vafa for pointing this connection out to us.}. This formulation was made more precise for AdS spaces as generalized distance conjecture which states that as $\Lambda$ goes to zero, there must be a tower of light states whose masses go to zero polynomialy fast in $\Lambda$ \cite{Lust:2019zwm}. If we replace $\Lambda$ with $\sim H^2$ and extend it to time-dependent backgrounds, we recover $m\lesssim H^c$ for some positive constant $c$.

Let us make the connection between our statement and the distance conjecture more precise. Consider a homogeneous background of an auxiliary canonically normalized massless field $\Phi$ that would reproduce the same action as the background we consider.
\begin{align}
    S=\int d^dx\sqrt{g} \frac{1}{2}(\partial_t\Phi)^2.
\end{align}
By matching the above action with the actual action $S=\int d^ds\sqrt{g}\mathcal{L}$ where $\mathcal{L}=\frac{1}{2\kappa^2}\mathcal{R}-V(\phi)-\frac{1}{2}(\partial_\mu\phi)^2$, we can define a formal notion of distance $\Delta\Phi$ for our rolling geometry that is given by $\int dt \sqrt{2\mathcal{L}}$. On the other hand, we have $\sqrt{\mathcal{L}}\propto H\propto t^{-1}$ as $t\rightarrow\infty$. Therefore, we find that this formal notion of distance between times $t_i$ and $t_f$ goes like $\propto \ln(t_f/t_i)$. If we apply the distance conjecture to this notion of distance, it would imply that there must be a tower of light states whose masses are exponential in $\ln(t/t_0)$, which would be polynomial in $t$. This is exactly what we find! Therefore, we find a derivation for generalized distance conjecture in scaling cosmologies. 

\section{Conclusions}

We showed that the most basic form of holography, that observables of quantum gravity live on the boundary of spacetime, has non-trivial implications for scalar field cosmologies. In particular, we showed that in a theory with asymptotic particles satisfying $m/H\rightarrow\infty$, which is generically expected \cite{Rudelius:2022gbz}, the expansion of the universe must satisfy TCC in the asymptotic future. Moreover, if the expansion goes like $t^p$, the masses of the weakly coupled particles must satisfy $m\lesssim t^{1-2p}$. We also showed that these conditions are deeply connected to the Swampland conjectures. In particular, the condition $p\leq1$ is equivalent to Trans-Planckian Censorship Conjecture in the future infinity, and the condition $m\lesssim t^{1-2p}$ matches with the generalized distance conjecture. Moreover, it provides an explicit bound for the coefficient of the distance conjecture that depends on the potential. However, the strongest constraint that we find is $m\lesssim\exp(-\phi/\sqrt{d-2})$ which is identical to the proposal of \cite{Etheredge:2022opl}. 

Our work presents a more fundamental explanation for various Swampland conjectures by connecting them to a more fundamental principle of quantum gravity, which is the holographic principle.

\section*{Acknowledgement}
We are grateful for countless productive discussions with Georges Obied that were essential to sharpening the core idea of this work. We thank Rashmish Mishra, Miguel Montero, John Stout, and Cumrun Vafa for helpful discussions. We also thank Georges Obied and Cumrun Vafa for valuable comments on the draft. The work of AB is supported by a grant from the Simons Foundation (602883, CV) and by the NSF grant PHY-2013858.
\appendix

\section{Scaling cosmologies}\label{A1}

In this Appendix we study FRW solutions that are driven by a scalar potential that exponentially decays at infinity. We show that in expansionary solution, exponential potentials lead to polynomial expansion at $t\rightarrow\infty$ and in contracting solutions, they lead to polynomial contraction at $t\rightarrow-\infty$. We also show that unless a smooth bounce happens, there is always a space-like singularity at finite past or future (i.e. big bang or big crunch).

We focus on FRW solution with zero spatial curvature. Let us start with the asymptotic future. We later use the time reversal symmetry of Einstein-Hilbert action to deduce similar conclusions about the past boundary. We assume that the evolution is driven by an potential that dies off exponentially in all directions of the field space. In that case, the vev of the scalar fields will always roll towards the infinity of the field space in the asymptotic future. 

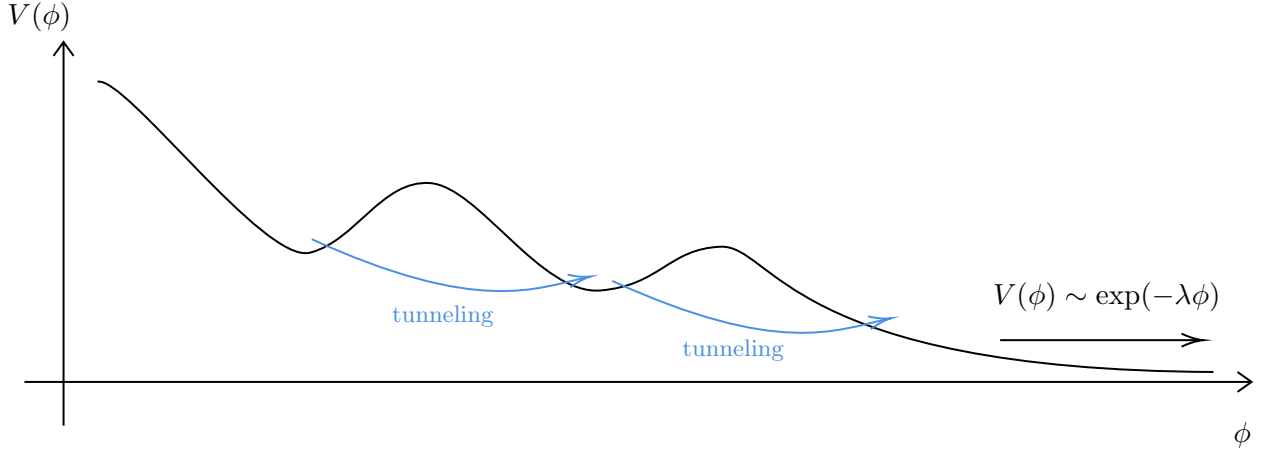
\begin{figure}[H]
    \centering

\tikzset{every picture/.style={line width=0.75pt}} 

\begin{tikzpicture}[x=0.75pt,y=0.75pt,yscale=-1,xscale=1]

\draw    (57.5,62) .. controls (71.5,59) and (140.58,153.87) .. (163.5,148) .. controls (186.42,142.13) and (199.32,111.23) .. (223.5,113) .. controls (247.68,114.77) and (278.66,169.49) .. (308.5,167) .. controls (338.34,164.51) and (342.5,146) .. (368.5,145) .. controls (394.5,144) and (401.5,207) .. (614.5,208) ;
\draw  (21,213.01) -- (633.5,213.01)(40.5,42) -- (40.5,235) (626.5,208.01) -- (633.5,213.01) -- (626.5,218.01) (35.5,49) -- (40.5,42) -- (45.5,49)  ;
\draw    (508,192) -- (607.5,192) ;
\draw [shift={(609.5,192)}, rotate = 180] [color={rgb, 255:red, 0; green, 0; blue, 0 }  ][line width=0.75]    (10.93,-3.29) .. controls (6.95,-1.4) and (3.31,-0.3) .. (0,0) .. controls (3.31,0.3) and (6.95,1.4) .. (10.93,3.29)   ;
\draw [color={rgb, 255:red, 74; green, 144; blue, 226 }  ,draw opacity=1 ]   (164.4,141.22) .. controls (247.27,179.22) and (276.09,166.16) .. (301.55,160.43) ;
\draw [shift={(303.5,160)}, rotate = 168.14] [color={rgb, 255:red, 74; green, 144; blue, 226 }  ,draw opacity=1 ][line width=0.75]    (10.93,-3.29) .. controls (6.95,-1.4) and (3.31,-0.3) .. (0,0) .. controls (3.31,0.3) and (6.95,1.4) .. (10.93,3.29)   ;
\draw [color={rgb, 255:red, 74; green, 144; blue, 226 }  ,draw opacity=1 ]   (314.4,162.22) .. controls (397.27,200.22) and (426.09,187.16) .. (451.55,181.43) ;
\draw [shift={(453.5,181)}, rotate = 168.14] [color={rgb, 255:red, 74; green, 144; blue, 226 }  ,draw opacity=1 ][line width=0.75]    (10.93,-3.29) .. controls (6.95,-1.4) and (3.31,-0.3) .. (0,0) .. controls (3.31,0.3) and (6.95,1.4) .. (10.93,3.29)   ;

\draw (503,161.4) node [anchor=north west][inner sep=0.75pt]    {$V( \phi ) \sim \exp( -\lambda \phi )$};
\draw (623,231.4) node [anchor=north west][inner sep=0.75pt]    {$\phi $};
\draw (11,20.4) node [anchor=north west][inner sep=0.75pt]    {$V( \phi )$};
\draw (203,173) node [anchor=north west][inner sep=0.75pt]  [color={rgb, 255:red, 74; green, 144; blue, 226 }  ,opacity=1 ] [align=left] {{\footnotesize tunneling}};
\draw (348,190) node [anchor=north west][inner sep=0.75pt]  [color={rgb, 255:red, 74; green, 144; blue, 226 }  ,opacity=1 ] [align=left] {{\footnotesize tunneling}};

\end{tikzpicture}
    \caption{If the potential always dies of exponentially, the universe cannot stay in any local minimum of the potential forever and it will eventually tunnel to a lower vacuum energy. Therefore, the at future infinity, the evolution of the universe is given by a scalar field rolling in an exponential potential.}
\end{figure}

Therefore, we just consider the quintessence solutions of the form $\phi\rightarrow \infty$ for an exponential potential $V(\phi)=V_0\exp(-\lambda\phi)$. We have two options: expanding universe or a contracting universe. As we will shortly see, the fate of these two universes is very different. In a contracting universe, the energy will blow up at finite time. Therefore, we will have a big crunch. While, in the expanding universe, the universe smoothly expands and dilutes. However, depending on the value of $\lambda$, the expansion can be accelerating or decelerating which changes the asymptotic boundary of spacetime. 

Let us start by setting up the equations of motion.
\begin{align}\label{FRWeom}
    &\frac{(d-1)(d-2)}{2}H^2=\frac{1}{2}\dot\phi^2+V(\phi)\nonumber\\
    &\ddot\phi+(d-1)H\dot\phi+V'(\phi)=0.
\end{align}
First thing to notice from the first equation is that $H$ can never vanish. Therefore, an expanding universe remains expanding and a contracting universe remains contracting\footnote{Note that here we are assuming there is no thermal background of other fields which is usually assumed in bounding cosmologies. We will come back to this assumption later.}. From the above equations one can also find
\begin{align}
    \dot H=-\frac{\dot\phi^2}{d-2}.
\end{align}
If we start with a contracting solution ($H<0$), then the absolute value of $H$ will keep increasing while $\phi$ rolls down the potential. Therefore, after some point, the Hubble energy must be mainly sourced by kinetic energy, which implies
\begin{align}
    \frac{(d-1)(d-2)}{2}H^2\simeq \frac{1}{2}\dot\phi^2\gg V(\phi).
\end{align}
From this we find that $V'$ is suppressed in the second equation of motion in \eqref{FRWeom}. So, we find, 
\begin{align}
    \ddot\phi\simeq-(d-1)H\dot\phi\simeq \sqrt\frac{d-1}{d-2}\dot\phi^2.
\end{align}
The solution to this equation is 
\begin{align}
    \phi(t)\simeq -\sqrt\frac{d-1}{d-2}\ln(c_1-\sqrt\frac{d-1}{d-2}\cdot t)+c_2,
\end{align}
where $c_1$ and $c_2$ are constants. As one can see, this solution diverges at finite time $t=c_1\sqrt\frac{d-2}{d-1}$. At this time, the kinetic energy diverges and drives the field to $\phi\rightarrow\infty$. This is a big crunch. Now let us look at the scale factor. We have 
\begin{align}
    \frac{\dot a}{a}=H\simeq -\frac{1}{\sqrt{(d-1)(d-2)}}\dot\phi=-\frac{1}{\sqrt{(d-1)(d-2)}(c_1-\sqrt\frac{d-1}{d-2}\cdot t}.
\end{align}
After integrating we find
\begin{align}\label{sfe}
    a\propto  (c_1-\sqrt\frac{d-1}{d-2}\cdot t)^\frac{1}{d-2}.
\end{align}
To see if the asymptotic boundary is spacelike or null, we should see if the contraction is slow enough to create a horizon\footnote{Fast expansions and slow contractions create cosmological horizons.}. In other words, if we take two points, can they communicate with each other before hitting the asymptotic boundary of spacetime? If any two points can, the asymptotic boundary is null, and if some points cannot, the asymptotic boundary is spacelike. 

The FRW metric is given by $ds^2=-dt^2+a(t)^2dX^2$. Therefore, the furthest coordinate distance that a signal can travel is $\int a^{-1}$. If this integral diverges, the asymptotic boundary is null. In the case of \eqref{sfe}, this integral clearly converges. Thus, the Penrose diragram looks like the following. 

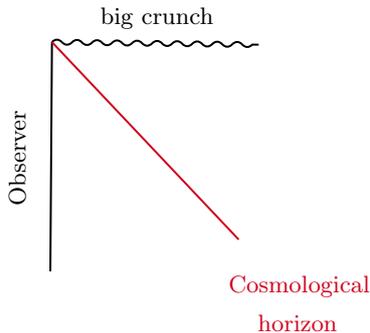
\begin{figure}[H]
    \centering

\tikzset{every picture/.style={line width=0.75pt}} 

\begin{tikzpicture}[x=0.75pt,y=0.75pt,yscale=-1,xscale=1]

\draw    (245.31,52.69) .. controls (247,51.04) and (248.66,51.06) .. (250.31,52.75) .. controls (251.96,54.44) and (253.62,54.46) .. (255.31,52.81) .. controls (257,51.17) and (258.66,51.19) .. (260.31,52.88) .. controls (261.96,54.57) and (263.62,54.59) .. (265.31,52.94) .. controls (267,51.29) and (268.66,51.31) .. (270.3,53) .. controls (271.95,54.69) and (273.61,54.71) .. (275.3,53.07) .. controls (276.99,51.42) and (278.65,51.44) .. (280.3,53.13) .. controls (281.95,54.82) and (283.61,54.84) .. (285.3,53.2) .. controls (286.99,51.55) and (288.65,51.57) .. (290.3,53.26) .. controls (291.95,54.95) and (293.61,54.97) .. (295.3,53.32) .. controls (296.99,51.68) and (298.65,51.7) .. (300.3,53.39) .. controls (301.95,55.08) and (303.61,55.1) .. (305.3,53.45) .. controls (306.99,51.8) and (308.65,51.82) .. (310.3,53.51) .. controls (311.95,55.2) and (313.61,55.22) .. (315.3,53.58) .. controls (316.99,51.93) and (318.65,51.95) .. (320.3,53.64) .. controls (321.95,55.33) and (323.61,55.35) .. (325.3,53.7) .. controls (326.99,52.06) and (328.65,52.08) .. (330.3,53.77) .. controls (331.95,55.46) and (333.61,55.48) .. (335.3,53.83) .. controls (336.99,52.19) and (338.65,52.21) .. (340.3,53.9) .. controls (341.95,55.59) and (343.61,55.61) .. (345.3,53.96) -- (348.5,54) -- (348.5,54) ;
\draw    (245.31,52.69) -- (244.5,168) ;
\draw [color={rgb, 255:red, 208; green, 2; blue, 27 }  ,draw opacity=1 ]   (245.31,52.69) -- (338.5,152) ;

\draw (222.5,136.5) node [anchor=north west][inner sep=0.75pt]  [font=\footnotesize,rotate=-270] [align=left] {\begin{minipage}[lt]{36.28pt}\setlength\topsep{0pt}
\begin{center}
Observer
\end{center}

\end{minipage}};
\draw (332,157) node [anchor=north west][inner sep=0.75pt]  [color={rgb, 255:red, 74; green, 144; blue, 226 }  ,opacity=1 ] [align=left] {\begin{minipage}[lt]{51.7pt}\setlength\topsep{0pt}
\begin{center}
{\footnotesize \textcolor[rgb]{0.82,0.01,0.11}{Cosmological}}\\{\footnotesize \textcolor[rgb]{0.82,0.01,0.11}{horizon}}
\end{center}

\end{minipage}};
\draw (268,34) node [anchor=north west][inner sep=0.75pt]  [font=\footnotesize] [align=left] {big crunch};

\end{tikzpicture}
    \caption{Penrose diagrams of contracting universes that are driven by an exponentially decaying scalar potential. The universe ends with a spacelike big-crunch.}
\end{figure}

In the above analysis we neglected creation of matter near the singularity. The inclusion of a thermal background would not have changed the qualitative behavior of the solution. To see this, note that in the above solution $H^2\propto a^{-2(d-2)}$. In dimensions greater than $3$ where gravity is dynamical, both matter and radiation increase slower than, or equal to, this rate. The energy density of a relativistic matter increases as $H_{rad}^2\propto a^{-d}$ and the energy density of a non-relativistic matter increases as $H_{matter}^2\propto a^{-(d-1)}$. Therefore, the qualitative behavior of the solution remains the same. 

Note that the asymptotic behavior of the solution at $t\rightarrow\infty$ is also independent from $\lambda$. So we find that, contracting solutions have a universal behavior at the asymptotic future and they end with a spacelike big crunch. 

Now let us consider the expanding solutions. In the expanding solutions, depending on exponent $\lambda$, the solution behaves qualitatively differently. For large enough $\lambda$, the expansion is so fast that the kinetic term dominates the Hubble energy. On the other hand, for small $\lambda$, all the terms are of the same order. 

\vspace{10pt}

\textbf{Case I: $\lambda<2\sqrt\frac{d-1}{d-2}$}

\vspace{10pt}

The attractor solution is
\begin{align}
    \phi=\frac{2}{\lambda}\ln(\sqrt\frac{\lambda V_0}{\frac{4(d-1)}{\lambda^2(d-2)}-1}t+c_1),
\end{align}
where $c_1$ is a constant. For this solution, all the terms in the Hubble energy are of the same order.
\begin{align}
    H^2\propto \dot\phi^2\propto V(\phi).
\end{align}
As $t\rightarrow\infty$ we have
\begin{align}
    \frac{\dot a}{a}\propto \frac{4}{(d-2)\lambda^2 t}.
\end{align}
Therefore, at future infinity, the scale factor goes like
\begin{align}
    a\propto t^p;~~~p=\frac{4}{(d-2)\lambda^2}.
\end{align}
\vspace{10pt}

\textbf{Case II: $\lambda>2\sqrt\frac{d-1}{d-2}$}

\vspace{10pt}
The attractor solution is
\begin{align}
    \phi=\sqrt{d-2}{d-1}\ln(t+c_1)+c_2,
\end{align}
where $c_1$ and $c_2$ are constants. For this solution, the kinetic term dominates and we have 
\begin{align}
    H^2\propto \dot\phi^2\gg V(\phi).
\end{align}
As $t\rightarrow\infty$ we have
\begin{align}
    \frac{\dot a}{a}\propto \frac{1}{(d-1)t}.
\end{align}
Therefore, at future infinity, the scale factor goes like
\begin{align}
    a\propto t^p;~~~p=\frac{1}{(d-1)}.
\end{align}

Now we study the shape of the Penrose diagram at future infinity. As we discussed earlier, the shape of the future boundary depends on the integral $\int a^{-1}$. For all the expanding solutions we found a polynomial dependence $a\sim t^p$. For $p>1$, the integral $\int a^{-1}$ converges which signals the existence of a spacelike boundary with a null cosmological horizon. While for $p<1$, there is no cosmological horizon and the future boundary must be null. 

Note that $p<1$ corresponds to $\lambda>\frac{2}{\sqrt{d-2}}$. Therefore, we find the following Penrose diagrams depending on the value of $\lambda$.

\begin{figure}[H]
    \centering

\tikzset{every picture/.style={line width=0.75pt}} 

\begin{tikzpicture}[x=0.75pt,y=0.75pt,yscale=-1,xscale=1]

\draw    (61.31,73.69) -- (164.5,75) ;
\draw    (61.31,73.69) -- (60.5,189) ;
\draw [color={rgb, 255:red, 208; green, 2; blue, 27 }  ,draw opacity=1 ]   (61.31,73.69) -- (154.5,173) ;
\draw    (411.31,81.69) -- (410.5,197) ;
\draw [color={rgb, 255:red, 0; green, 0; blue, 0 }  ,draw opacity=1 ]   (411.31,81.69) -- (504.5,181) ;

\draw (122,51.4) node [anchor=north west][inner sep=0.75pt]    {$\mathscr{I} ^{+}$};
\draw (38.5,157.5) node [anchor=north west][inner sep=0.75pt]  [font=\footnotesize,rotate=-270] [align=left] {\begin{minipage}[lt]{36.28pt}\setlength\topsep{0pt}
\begin{center}
Observer
\end{center}

\end{minipage}};
\draw (148,178) node [anchor=north west][inner sep=0.75pt]  [color={rgb, 255:red, 74; green, 144; blue, 226 }  ,opacity=1 ] [align=left] {\begin{minipage}[lt]{51.7pt}\setlength\topsep{0pt}
\begin{center}
{\footnotesize \textcolor[rgb]{0.82,0.01,0.11}{Cosmological}}\\{\footnotesize \textcolor[rgb]{0.82,0.01,0.11}{horizon}}
\end{center}

\end{minipage}};
\draw (171,68) node [anchor=north west][inner sep=0.75pt]  [font=\footnotesize] [align=left] {\begin{minipage}[lt]{65.76pt}\setlength\topsep{0pt}
\begin{center}
Asymptotic future
\end{center}

\end{minipage}};
\draw (443,88.4) node [anchor=north west][inner sep=0.75pt]    {$\mathscr{I} ^{+}$};
\draw (388.5,165.5) node [anchor=north west][inner sep=0.75pt]  [font=\footnotesize,rotate=-270] [align=left] {\begin{minipage}[lt]{36.28pt}\setlength\topsep{0pt}
\begin{center}
Observer
\end{center}

\end{minipage}};
\draw (480,136) node [anchor=north west][inner sep=0.75pt]  [font=\footnotesize] [align=left] {\begin{minipage}[lt]{65.76pt}\setlength\topsep{0pt}
\begin{center}
Asymptotic future
\end{center}

\end{minipage}};
\draw (109,250.4) node [anchor=north west][inner sep=0.75pt]  [font=\scriptsize]  {$\lambda < \frac{2}{\sqrt{d-2}}$};
\draw (430,251.4) node [anchor=north west][inner sep=0.75pt]  [font=\scriptsize]  {$\lambda  >\frac{2}{\sqrt{d-2}}$};

\end{tikzpicture}
    \caption{Depending on the value of $\lambda$, the Penrose diagram takes a different shape in $t\rightarrow\infty$. For $\lambda>\frac{2}{\sqrt{d-2}}$, the future infinity is decelerating and the asymptotic future $\mathscr{I}^+$ is null. However, for $\lambda<\frac{2}{\sqrt{d-2}}$, the asymptotic future has an accelerating expansion which creates a cosmological horizon.}
\end{figure}
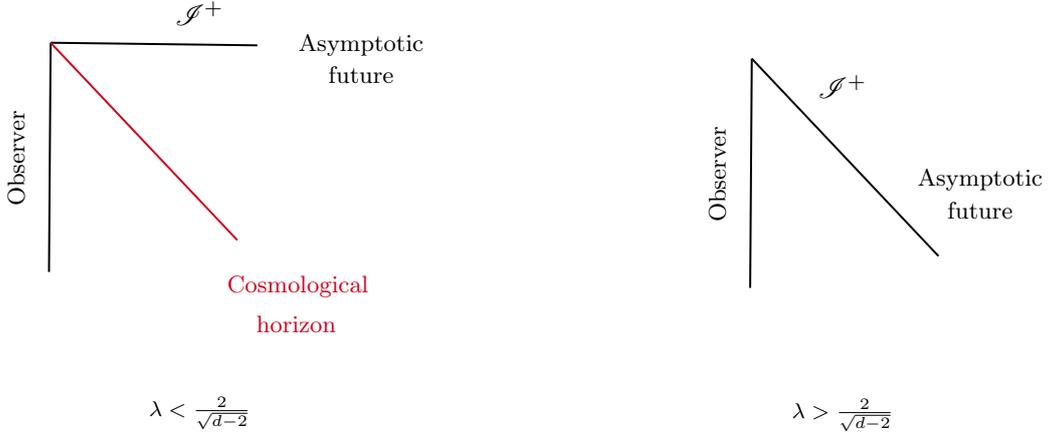

Now, we can time reverse the above analysis and make similar statements about the past infinity of the expanding and contracting universes. 

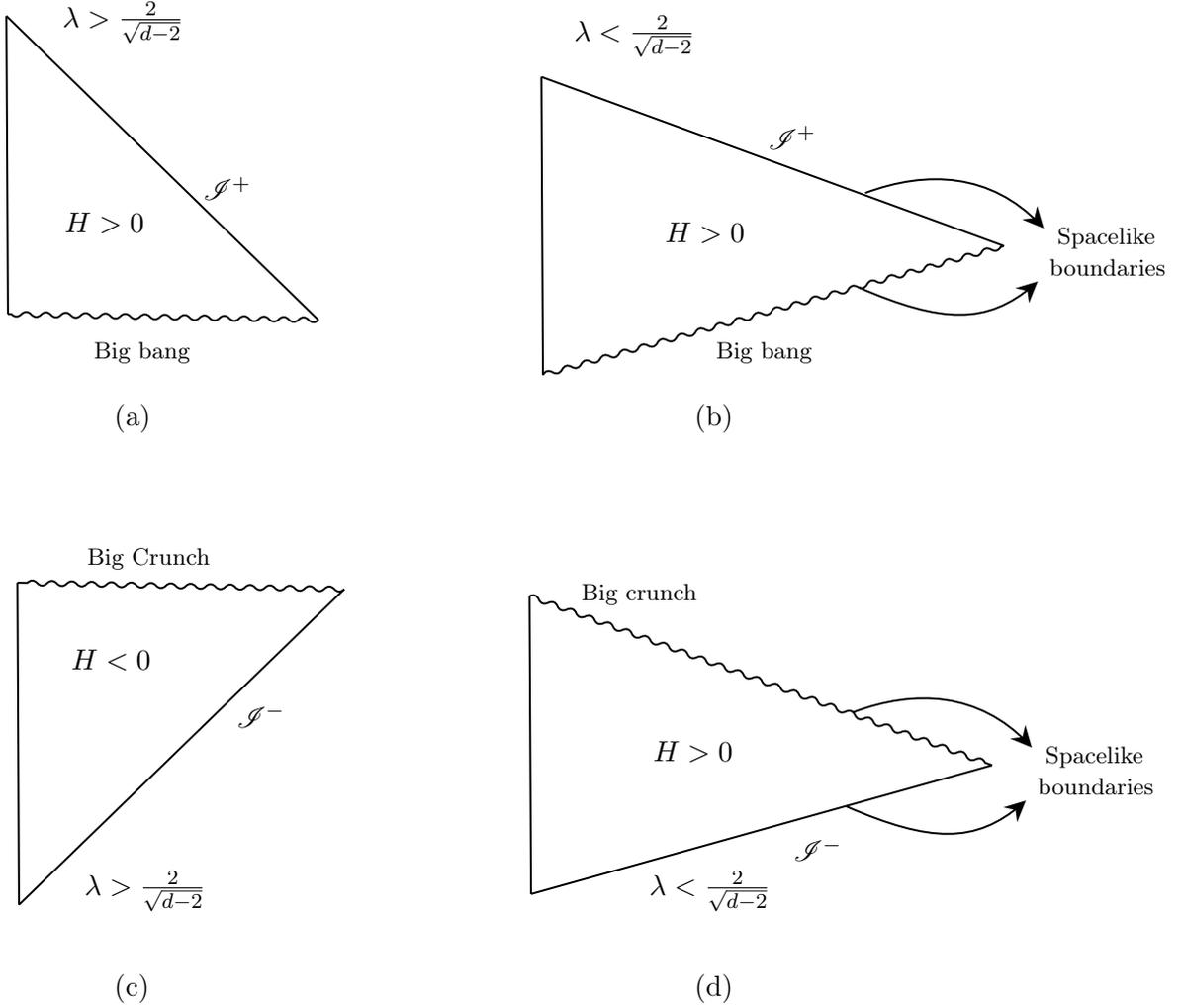
\begin{figure}[H]
    \centering

\tikzset{every picture/.style={line width=0.75pt}} 

\begin{tikzpicture}[x=0.75pt,y=0.75pt,yscale=-1,xscale=1]

\draw    (47,67) -- (204.5,221.62) ;
\draw    (47,67) -- (47.79,218.53) ;
\draw    (204.5,221.62) .. controls (202.8,223.25) and (201.13,223.22) .. (199.5,221.52) .. controls (197.87,219.82) and (196.2,219.79) .. (194.5,221.42) .. controls (192.81,223.05) and (191.14,223.02) .. (189.5,221.33) .. controls (187.87,219.63) and (186.2,219.6) .. (184.5,221.23) .. controls (182.8,222.86) and (181.13,222.83) .. (179.5,221.13) .. controls (177.87,219.43) and (176.21,219.4) .. (174.51,221.03) .. controls (172.81,222.66) and (171.14,222.63) .. (169.51,220.93) .. controls (167.88,219.23) and (166.21,219.2) .. (164.51,220.83) .. controls (162.81,222.46) and (161.14,222.43) .. (159.51,220.73) .. controls (157.87,219.04) and (156.2,219.01) .. (154.51,220.64) .. controls (152.81,222.27) and (151.14,222.24) .. (149.51,220.54) .. controls (147.88,218.84) and (146.21,218.81) .. (144.51,220.44) .. controls (142.81,222.07) and (141.14,222.04) .. (139.51,220.34) .. controls (137.88,218.64) and (136.21,218.61) .. (134.51,220.24) .. controls (132.81,221.87) and (131.14,221.84) .. (129.51,220.14) .. controls (127.88,218.44) and (126.22,218.41) .. (124.52,220.04) .. controls (122.82,221.67) and (121.15,221.64) .. (119.52,219.94) .. controls (117.88,218.25) and (116.21,218.22) .. (114.52,219.85) .. controls (112.82,221.48) and (111.15,221.45) .. (109.52,219.75) .. controls (107.89,218.05) and (106.22,218.02) .. (104.52,219.65) .. controls (102.82,221.28) and (101.15,221.25) .. (99.52,219.55) .. controls (97.89,217.85) and (96.22,217.82) .. (94.52,219.45) .. controls (92.82,221.08) and (91.15,221.05) .. (89.52,219.35) .. controls (87.89,217.65) and (86.22,217.62) .. (84.52,219.25) .. controls (82.83,220.88) and (81.16,220.85) .. (79.52,219.16) .. controls (77.89,217.46) and (76.23,217.43) .. (74.53,219.06) .. controls (72.83,220.69) and (71.16,220.66) .. (69.53,218.96) .. controls (67.9,217.26) and (66.23,217.23) .. (64.53,218.86) .. controls (62.83,220.49) and (61.16,220.46) .. (59.53,218.76) .. controls (57.9,217.06) and (56.23,217.03) .. (54.53,218.66) .. controls (52.83,220.29) and (51.16,220.26) .. (49.53,218.56) -- (47.79,218.53) -- (47.79,218.53) ;
\draw    (317,98) -- (317.79,249.53) ;
\draw    (317,98) -- (550.5,184) ;
\draw    (317.79,249.53) .. controls (318.94,247.47) and (320.54,247.02) .. (322.6,248.17) .. controls (324.65,249.32) and (326.26,248.87) .. (327.41,246.82) .. controls (328.56,244.76) and (330.17,244.31) .. (332.23,245.46) .. controls (334.28,246.61) and (335.89,246.16) .. (337.04,244.11) .. controls (338.19,242.05) and (339.79,241.6) .. (341.85,242.75) .. controls (343.9,243.9) and (345.51,243.45) .. (346.66,241.4) .. controls (347.81,239.34) and (349.42,238.89) .. (351.48,240.04) .. controls (353.53,241.19) and (355.14,240.74) .. (356.29,238.69) .. controls (357.44,236.63) and (359.04,236.18) .. (361.1,237.33) .. controls (363.15,238.48) and (364.76,238.03) .. (365.92,235.98) .. controls (367.07,233.92) and (368.67,233.47) .. (370.73,234.62) .. controls (372.78,235.77) and (374.39,235.32) .. (375.54,233.27) .. controls (376.69,231.21) and (378.29,230.76) .. (380.35,231.91) .. controls (382.4,233.06) and (384.01,232.61) .. (385.17,230.56) .. controls (386.32,228.5) and (387.92,228.05) .. (389.98,229.2) .. controls (392.03,230.35) and (393.64,229.9) .. (394.79,227.85) .. controls (395.94,225.79) and (397.55,225.34) .. (399.61,226.49) .. controls (401.66,227.64) and (403.27,227.19) .. (404.42,225.14) .. controls (405.57,223.08) and (407.17,222.63) .. (409.23,223.78) .. controls (411.29,224.93) and (412.89,224.48) .. (414.04,222.42) .. controls (415.2,220.37) and (416.81,219.92) .. (418.86,221.07) .. controls (420.92,222.22) and (422.52,221.77) .. (423.67,219.71) .. controls (424.82,217.66) and (426.43,217.21) .. (428.48,218.36) .. controls (430.54,219.51) and (432.15,219.06) .. (433.3,217) .. controls (434.45,214.95) and (436.06,214.5) .. (438.11,215.65) .. controls (440.17,216.8) and (441.77,216.35) .. (442.92,214.29) .. controls (444.07,212.24) and (445.68,211.79) .. (447.73,212.94) .. controls (449.79,214.09) and (451.4,213.64) .. (452.55,211.58) .. controls (453.7,209.53) and (455.31,209.08) .. (457.36,210.23) .. controls (459.42,211.38) and (461.02,210.93) .. (462.17,208.87) .. controls (463.33,206.82) and (464.94,206.37) .. (466.99,207.52) .. controls (469.05,208.67) and (470.65,208.22) .. (471.8,206.16) .. controls (472.95,204.11) and (474.56,203.66) .. (476.61,204.81) .. controls (478.67,205.96) and (480.27,205.51) .. (481.42,203.45) .. controls (482.58,201.4) and (484.19,200.95) .. (486.24,202.1) .. controls (488.3,203.25) and (489.9,202.8) .. (491.05,200.74) .. controls (492.2,198.69) and (493.81,198.24) .. (495.86,199.39) .. controls (497.92,200.54) and (499.52,200.09) .. (500.67,198.03) .. controls (501.82,195.97) and (503.43,195.52) .. (505.49,196.67) .. controls (507.54,197.82) and (509.15,197.37) .. (510.3,195.32) .. controls (511.45,193.26) and (513.05,192.81) .. (515.11,193.96) .. controls (517.16,195.11) and (518.77,194.66) .. (519.93,192.61) .. controls (521.08,190.55) and (522.68,190.1) .. (524.74,191.25) .. controls (526.79,192.4) and (528.4,191.95) .. (529.55,189.9) .. controls (530.7,187.84) and (532.3,187.39) .. (534.36,188.54) .. controls (536.41,189.69) and (538.02,189.24) .. (539.18,187.19) .. controls (540.33,185.13) and (541.93,184.68) .. (543.99,185.83) .. controls (546.04,186.98) and (547.65,186.53) .. (548.8,184.48) -- (550.5,184) -- (550.5,184) ;
\draw    (480.5,157) .. controls (511.7,145.3) and (543.85,146.91) .. (568.61,172.95) ;
\draw [shift={(570.5,175)}, rotate = 228.24] [fill={rgb, 255:red, 0; green, 0; blue, 0 }  ][line width=0.08]  [draw opacity=0] (10.72,-5.15) -- (0,0) -- (10.72,5.15) -- (7.12,0) -- cycle    ;
\draw    (477,205) .. controls (519.2,223.43) and (542.1,223.98) .. (565.34,203.92) ;
\draw [shift={(567.5,202)}, rotate = 137.49] [fill={rgb, 255:red, 0; green, 0; blue, 0 }  ][line width=0.08]  [draw opacity=0] (10.72,-5.15) -- (0,0) -- (10.72,5.15) -- (7.12,0) -- cycle    ;
\draw    (53.33,519) -- (217.5,358.48) ;
\draw    (53.33,519) -- (52.56,355.13) ;
\draw    (217.5,358.48) .. controls (215.8,360.11) and (214.13,360.08) .. (212.5,358.38) .. controls (210.87,356.68) and (209.2,356.65) .. (207.5,358.28) .. controls (205.8,359.91) and (204.13,359.88) .. (202.5,358.18) .. controls (200.87,356.48) and (199.2,356.45) .. (197.5,358.08) .. controls (195.8,359.71) and (194.14,359.67) .. (192.51,357.97) .. controls (190.88,356.27) and (189.21,356.24) .. (187.51,357.87) .. controls (185.81,359.5) and (184.14,359.47) .. (182.51,357.77) .. controls (180.88,356.07) and (179.21,356.04) .. (177.51,357.67) .. controls (175.81,359.3) and (174.14,359.27) .. (172.51,357.57) .. controls (170.88,355.87) and (169.21,355.84) .. (167.51,357.47) .. controls (165.81,359.1) and (164.14,359.06) .. (162.51,357.36) .. controls (160.88,355.66) and (159.21,355.63) .. (157.51,357.26) .. controls (155.81,358.89) and (154.14,358.86) .. (152.51,357.16) .. controls (150.88,355.46) and (149.21,355.43) .. (147.51,357.06) .. controls (145.81,358.69) and (144.15,358.66) .. (142.52,356.96) .. controls (140.89,355.26) and (139.22,355.22) .. (137.52,356.85) .. controls (135.82,358.48) and (134.15,358.45) .. (132.52,356.75) .. controls (130.89,355.05) and (129.22,355.02) .. (127.52,356.65) .. controls (125.82,358.28) and (124.15,358.25) .. (122.52,356.55) .. controls (120.89,354.85) and (119.22,354.82) .. (117.52,356.45) .. controls (115.82,358.08) and (114.15,358.05) .. (112.52,356.35) .. controls (110.89,354.65) and (109.22,354.61) .. (107.52,356.24) .. controls (105.82,357.87) and (104.15,357.84) .. (102.52,356.14) .. controls (100.89,354.44) and (99.22,354.41) .. (97.52,356.04) .. controls (95.82,357.67) and (94.16,357.64) .. (92.53,355.94) .. controls (90.9,354.24) and (89.23,354.21) .. (87.53,355.84) .. controls (85.83,357.47) and (84.16,357.44) .. (82.53,355.74) .. controls (80.9,354.04) and (79.23,354) .. (77.53,355.63) .. controls (75.83,357.26) and (74.16,357.23) .. (72.53,355.53) .. controls (70.9,353.83) and (69.23,353.8) .. (67.53,355.43) .. controls (65.83,357.06) and (64.16,357.03) .. (62.53,355.33) .. controls (60.9,353.63) and (59.23,353.6) .. (57.53,355.23) -- (52.56,355.13) -- (52.56,355.13) ;
\draw    (311,362) -- (311.79,513.53) ;
\draw    (311,362) .. controls (313.14,361.01) and (314.7,361.59) .. (315.69,363.73) .. controls (316.68,365.87) and (318.24,366.45) .. (320.38,365.46) .. controls (322.52,364.47) and (324.09,365.04) .. (325.08,367.18) .. controls (326.07,369.32) and (327.63,369.9) .. (329.77,368.91) .. controls (331.91,367.92) and (333.47,368.5) .. (334.46,370.64) .. controls (335.45,372.78) and (337.01,373.36) .. (339.15,372.37) .. controls (341.29,371.38) and (342.85,371.96) .. (343.84,374.1) .. controls (344.83,376.24) and (346.4,376.81) .. (348.54,375.82) .. controls (350.68,374.83) and (352.24,375.41) .. (353.23,377.55) .. controls (354.22,379.69) and (355.78,380.27) .. (357.92,379.28) .. controls (360.06,378.29) and (361.62,378.87) .. (362.61,381.01) .. controls (363.6,383.15) and (365.16,383.73) .. (367.3,382.74) .. controls (369.44,381.75) and (371,382.32) .. (371.99,384.46) .. controls (372.98,386.6) and (374.55,387.18) .. (376.69,386.19) .. controls (378.83,385.2) and (380.39,385.78) .. (381.38,387.92) .. controls (382.37,390.06) and (383.93,390.64) .. (386.07,389.65) .. controls (388.21,388.66) and (389.77,389.24) .. (390.76,391.38) .. controls (391.75,393.52) and (393.31,394.1) .. (395.45,393.11) .. controls (397.59,392.12) and (399.16,392.69) .. (400.15,394.83) .. controls (401.14,396.97) and (402.7,397.55) .. (404.84,396.56) .. controls (406.98,395.57) and (408.54,396.15) .. (409.53,398.29) .. controls (410.52,400.43) and (412.08,401.01) .. (414.22,400.02) .. controls (416.36,399.03) and (417.92,399.61) .. (418.91,401.75) .. controls (419.9,403.89) and (421.47,404.46) .. (423.61,403.47) .. controls (425.75,402.48) and (427.31,403.06) .. (428.3,405.2) .. controls (429.29,407.34) and (430.85,407.92) .. (432.99,406.93) .. controls (435.13,405.94) and (436.69,406.52) .. (437.68,408.66) .. controls (438.67,410.8) and (440.23,411.38) .. (442.37,410.39) .. controls (444.51,409.4) and (446.07,409.97) .. (447.06,412.11) .. controls (448.05,414.25) and (449.62,414.83) .. (451.76,413.84) .. controls (453.9,412.85) and (455.46,413.43) .. (456.45,415.57) .. controls (457.44,417.71) and (459,418.29) .. (461.14,417.3) .. controls (463.28,416.31) and (464.84,416.89) .. (465.83,419.03) .. controls (466.82,421.17) and (468.38,421.74) .. (470.52,420.75) .. controls (472.66,419.76) and (474.23,420.34) .. (475.22,422.48) .. controls (476.21,424.62) and (477.77,425.2) .. (479.91,424.21) .. controls (482.05,423.22) and (483.61,423.8) .. (484.6,425.94) .. controls (485.59,428.08) and (487.15,428.66) .. (489.29,427.67) .. controls (491.43,426.68) and (492.99,427.25) .. (493.98,429.39) .. controls (494.97,431.53) and (496.54,432.11) .. (498.68,431.12) .. controls (500.82,430.13) and (502.38,430.71) .. (503.37,432.85) .. controls (504.36,434.99) and (505.92,435.57) .. (508.06,434.58) .. controls (510.2,433.59) and (511.76,434.17) .. (512.75,436.31) .. controls (513.74,438.45) and (515.3,439.02) .. (517.44,438.03) .. controls (519.58,437.04) and (521.14,437.62) .. (522.13,439.76) .. controls (523.12,441.9) and (524.69,442.48) .. (526.83,441.49) .. controls (528.97,440.5) and (530.53,441.08) .. (531.52,443.22) .. controls (532.51,445.36) and (534.07,445.94) .. (536.21,444.95) .. controls (538.35,443.96) and (539.91,444.53) .. (540.9,446.67) -- (544.5,448) -- (544.5,448) ;
\draw    (311.79,513.53) -- (544.5,448) ;
\draw    (474.5,421) .. controls (505.7,409.3) and (537.85,410.91) .. (562.61,436.95) ;
\draw [shift={(564.5,439)}, rotate = 228.24] [fill={rgb, 255:red, 0; green, 0; blue, 0 }  ][line width=0.08]  [draw opacity=0] (10.72,-5.15) -- (0,0) -- (10.72,5.15) -- (7.12,0) -- cycle    ;
\draw    (471,469) .. controls (513.2,487.43) and (536.1,487.98) .. (559.34,467.92) ;
\draw [shift={(561.5,466)}, rotate = 137.49] [fill={rgb, 255:red, 0; green, 0; blue, 0 }  ][line width=0.08]  [draw opacity=0] (10.72,-5.15) -- (0,0) -- (10.72,5.15) -- (7.12,0) -- cycle    ;

\draw (89.73,231.36) node [anchor=north west][inner sep=0.75pt]  [font=\footnotesize] [align=left] {Big bang};
\draw (74,58.4) node [anchor=north west][inner sep=0.75pt]    {$\lambda  >\frac{2}{\sqrt{d-2}}$};
\draw (403.73,231.36) node [anchor=north west][inner sep=0.75pt]  [font=\footnotesize] [align=left] {Big bang};
\draw (332,65.4) node [anchor=north west][inner sep=0.75pt]    {$\lambda < \frac{2}{\sqrt{d-2}}$};
\draw (75,165.4) node [anchor=north west][inner sep=0.75pt]    {$H >0$};
\draw (378,170.4) node [anchor=north west][inner sep=0.75pt]    {$H >0$};
\draw (572,173) node [anchor=north west][inner sep=0.75pt]  [font=\footnotesize] [align=left] {\begin{minipage}[lt]{43.1pt}\setlength\topsep{0pt}
\begin{center}
Spacelike \\boundaries
\end{center}

\end{minipage}};
\draw (145,147.4) node [anchor=north west][inner sep=0.75pt]    {$\mathscr{I} ^{+}$};
\draw (430,121.4) node [anchor=north west][inner sep=0.75pt]    {$\mathscr{I}^{+}$};
\draw (86.44,336.04) node [anchor=north west][inner sep=0.75pt]  [font=\footnotesize] [align=left] {Big Crunch};
\draw (85,500) node [anchor=north west][inner sep=0.75pt]    {$\lambda  >\frac{2}{\sqrt{d-2}}$};
\draw (335.73,354.36) node [anchor=north west][inner sep=0.75pt]  [font=\footnotesize] [align=left] {Big crunch};
\draw (370,500) node [anchor=north west][inner sep=0.75pt]    {$\lambda < \frac{2}{\sqrt{d-2}}$};
\draw (78.38,387.98) node [anchor=north west][inner sep=0.75pt]    {$H< 0$};
\draw (372,434.4) node [anchor=north west][inner sep=0.75pt]    {$H >0$};
\draw (566,437) node [anchor=north west][inner sep=0.75pt]  [font=\footnotesize] [align=left] {\begin{minipage}[lt]{43.1pt}\setlength\topsep{0pt}
\begin{center}
Spacelike \\boundaries
\end{center}

\end{minipage}};
\draw (162,415.4) node [anchor=north west][inner sep=0.75pt]    {$\mathscr{I} ^{-}$};
\draw (443,483.4) node [anchor=north west][inner sep=0.75pt]    {$\mathscr{I} ^{-}$};
\draw (100,263) node [anchor=north west][inner sep=0.75pt]   [align=left] {(a)};
\draw (393,263) node [anchor=north west][inner sep=0.75pt]   [align=left] {(b)};
\draw (100,553) node [anchor=north west][inner sep=0.75pt]   [align=left] {(c)};
\draw (393,553) node [anchor=north west][inner sep=0.75pt]   [align=left] {(d)};

\end{tikzpicture}
    \caption{Penrose diagrams of: (a) expanding universe with deccelerating asymptotic future, (b) expanding universe with accelerating asymptotic future, (c) contracting universe with deccelerating asymptotic past, and (d) contracting universe with accelerating asymptotic past.}
    \label{FOU}
\end{figure}

We find that the past infinity of the expanding universes is always spacelike and the past infinity of contracting universes depends on the coefficient $\lambda$ that drives the evolution. Therefore, we find the following four options (Fig. \ref{FOU}).

\bibliographystyle{unsrt}
\bibliography{References}
\end{document}